\documentclass[preprint2]{aastex61}
\usepackage{amsmath}
\usepackage{graphicx}
\usepackage[normalem]{ulem}
\usepackage{color}

\shorttitle{Simulated Encounters of PSP with a CH Jet}
\shortauthors{Roberts et al.}

\begin{document}
\title{Simulated Encounters of Parker Solar Probe with a Coronal-Hole Jet}
\author{Merrill A Roberts and Vadim M Uritsky}
\affil{The Catholic University of America, 620 Michigan Ave. NE, Washington, DC, 20061}
\affil{Heliophysics Science Division, NASA Goddard Space Flight Center, Greenbelt, MD 20771, USA}
\author{C Richard DeVore and Judith T Karpen}
\affil{Heliophysics Science Division, NASA Goddard Space Flight Center, Greenbelt, MD 20771, USA}

\begin{abstract}
{Solar coronal jets are small, transient, collimated ejections most easily observed in coronal holes (CHs).  The upcoming Parker Solar Probe (PSP) mission provides the first opportunity to encounter CH jets \textit{in situ} near the Sun and examine their internal structure and dynamics.  Using projected mission orbital parameters, we have simulated PSP encounters with a fully three-dimensional magnetohydrodynamic (MHD) model of a CH jet.  We find that three internal jet regions, featuring different wave modes and levels of compressibility, have distinct identifying signatures detectable by PSP.  The leading Alfv\'{e}n wave front and its immediate wake are characterized by trans-Alfv\'{e}nic plasma flows with mild density enhancements.  This front exhibits characteristics of a fast switch-on MHD shock, whose arrival is signaled by the sudden onset of large-amplitude transverse velocity and magnetic-field oscillations highly correlated in space and time.  The trailing portion is characterized by supersonic but sub-Alfv\'{e}nic outflows of dense plasma with uncorrelated velocity and magnetic-field oscillations.  This compressible region contains most of the jet's mass.  The volume between the immediate wake and dense jet, the remote wake, mixes and transitions the characteristics of the two other regions.  In addition to probing each region separately, we also simulate a co-rotational PSP-jet encounter.  In this scenario, the simulated spacecraft hovers over the jet-producing CH, as may occur during the mission's co-rotational phases, sampling each jet region in turn.  We estimate that PSP will encounter numerous CH jets over the lifetime of the mission.}

\end{abstract}

\newpage

\section{Introduction}
\label{sec:intro}

\begin{figure*}
\begin{center}
\includegraphics[width=18cm]{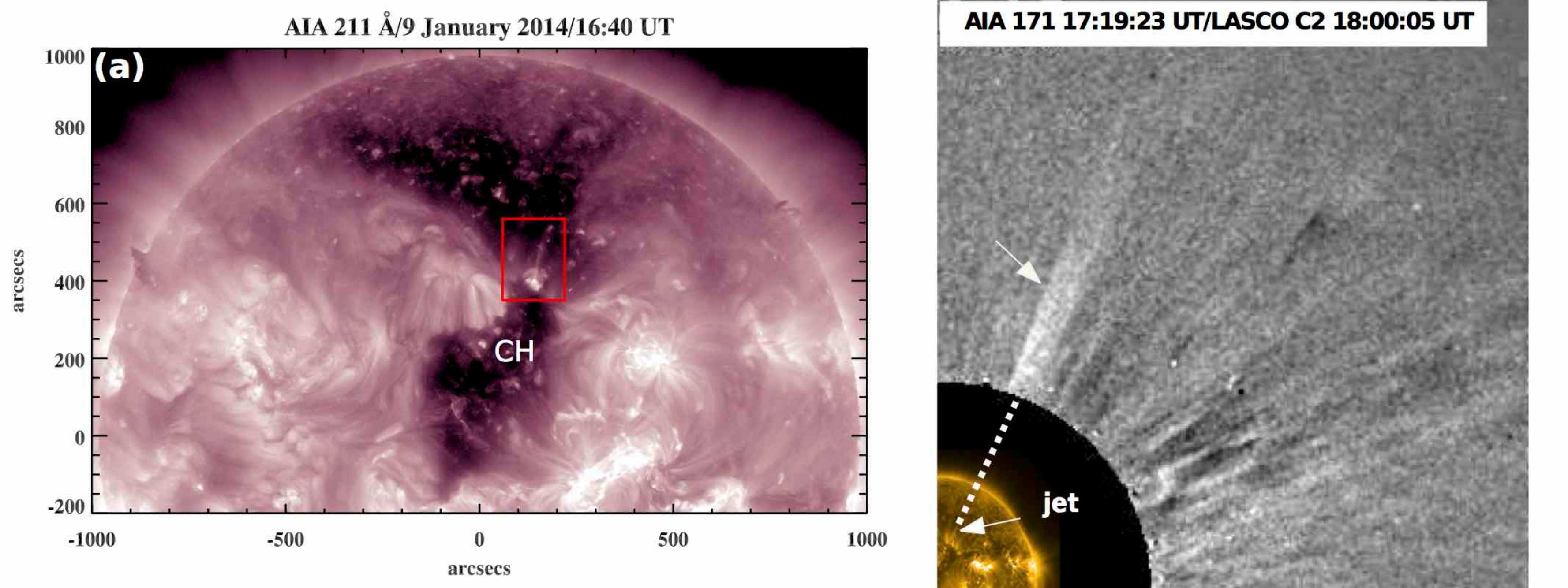}
\caption{{SDO/AIA and SOHO/LASCO images of a coronal-hole jet that occurred on 2014 January 9. Left panel: AIA image of the equatorial CH and jet source region (within the red box) . Right panel: Composite image of the jet in EUV (AIA 171\AA) and white light (LASCO C2). From \citet{kumar2018} }}
\label{fig:wsjet}
\end{center}
\end{figure*}

Solar coronal jets \citep[and references therein]{raouafi2016} are small-scale, transient, highly collimated, impulsive flows of plasma that have been observed in all regions of the Sun.  Due to the lower ambient emission from the cooler, low-density plasma prevalent in coronal holes, the Extreme Ultraviolet (EUV) and X-ray signatures of jets are more enhanced above background and therefore are more frequently observed in those regions. Coronal-hole (CH) jets typically exhibit structures with a wide base and a tall spire or curtain of bright material (see Figure \ref{fig:wsjet}) in an `Eiffel tower' or `$\lambda$' geometry \citep[see, e.g.,][]{nistico09}. Many CH jets have been observed by white-light coronagraphs to propagate into the inner heliosphere \citep[e.g.,][]{wang1998b,wang2002,kumar2018}, and it has been suggested \citep{neugebauer2012} that they may be the origin of microstreams detected \textit{in situ} in the solar wind \citep{neugebauer1995}.  Observations have also detected distinctive fast-moving helical structures in some CH jets, which have been identified as radially outflowing nonlinear Alfv\'{e}n waves \citep{cirtain2007}.  As nonlinear Alfv\'{e}n waves have been detected in the interplanetary medium \citep{gosling10,marubashi10}, it is plausible that CH jets may be the source of some of these disturbances as well. 

Due to the small scale and transient nature of CH jets, as well as the reliance on remote-sensing measurements, many aspects of these coronal events remain not well understood.  Among them are the jet's interior structure and dynamics, and the extent to which the jet maintains its coherence as it propagates into the heliosphere.  Our goal is to examine these properties as predicted by a high-resolution magnetohydrodynamic (MHD) model of a CH jet, in order to make testable predictions for the upcoming Parker Solar Probe (PSP) mission.

In recent work, \citet[][hereafter K17]{karpen17} simulated a CH jet driven by magnetic reconnection following onset of a kink-like instability within the closed field of an embedded bipole in a coronal hole. The model included spherical geometry, solar gravity, and an isothermal wind, and simulated the propagation of the Alfv\'{e}n wave front into the outer corona to about 5$R_\sun$ (where $R_\odot\approx7\times10^5$ km is the solar radius). The underlying kink-instability mechanism was identified and the resultant nonlinear, helical, Alfv\'{e}nic jets were analyzed extensively in Cartesian, gravity-free environments by \citet{pariat09,pariat10,pariat15,pariat16}.

Subsequently, \citet[][hereafter U17]{uritsky17} investigated the internal structure and turbulent dynamics of the modeled jet and determined that there are three distinct regions within the propagating structure: (1) The immediate wake, a region of incompressible flow moving at a speed comparable to the local Alfv\'{e}n speed and located behind the leading edge of the jet; (2) The dense jet, a region of compressible flow and overdense plasma moving at a much lower, but still supersonic, speed which forms closer to the solar surface; (3) The remote wake, an intervening transition region where the characteristics of the other two regimes are mixed together.  This investigation showed that the model produced CH jets with complex internal dynamics and structure that were preserved as the jet propagated into the outer corona.  Quite recently, \cite{horbury2018} further suggested that large-amplitude velocity enhancements detected by the HELIOS spacecraft may be examples of the Alfv\'{e}nic outflows reported by K17 and U17, providing further impetus for our exploration of {\it{in situ}} signatures of CH jets that may be observed with PSP.

Parker Solar Probe \citep[previously named Solar Probe Plus;][]{fox16} is scheduled to launch in August 2018. PSP will obtain the first \textit{in situ} measurements of the Sun closer than 0.3 AU, eventually approaching to less than 10$R_\sun$. The minimum periapsis of the spacecraft is currently planned at $6.16 \times 10^6$ km ($8.86 R_\sun$), and its much-anticipated trip into the outer corona is expected to return a wealth of information.  With PSP projected to reach its minimum perihelion near the next predicted solar maximum in December 2024, it is likely to encounter equatorial coronal holes and the ubiquitous CH jets that they contain. This raises the question: What observational signatures should be anticipated when PSP encounters a CH jet?

We have addressed this question by extending the CH jet model investigated by K17 and U17 beyond the previous 5$R_\sun$ limit, into the inner heliosphere that will be probed by PSP.  The results of this new MHD simulation, whose model space extends radially to 60$R_\sun$, demonstrate that CH jets could propagate coherently into the predicted orbital range of PSP.  Within this physical model, we present the first virtual fly-throughs of PSP encountering a CH jet with detailed internal structure and dynamics.  We use the PSP spacecraft trajectory to guide a spatio-temporal analysis of the jet in four dimensions and to develop predictions of the resulting jet signatures. We simulate PSP-jet encounters that probe each of the three regions described by U17, as well as a co-rotational encounter that probes the entire extent of the propagating jet, and extract the specific signatures expected for each type of encounter.  We also estimate the frequency of occurrence of PSP encounters with CH jets.

A brief description of the jet model is given in \S \ref{sec:model}.  We discuss the PSP mission in \S \ref{sec:psp}, and we describe the projected PSP orbital parameters and their interpolation into our model space in \S \ref{sec:ftm}.  The results of the four simulated fly-through trajectories are presented in \S \ref{sec:results}.  Our estimate for the frequency of PSP/jet encounters is given in \S \ref{sec:freq}.  We summarize the implications of our findings and discuss future research in \S \ref{sec:summary}.

\section{Model Description}
\label{sec:model}

\begin{figure}
\includegraphics[width=\linewidth]{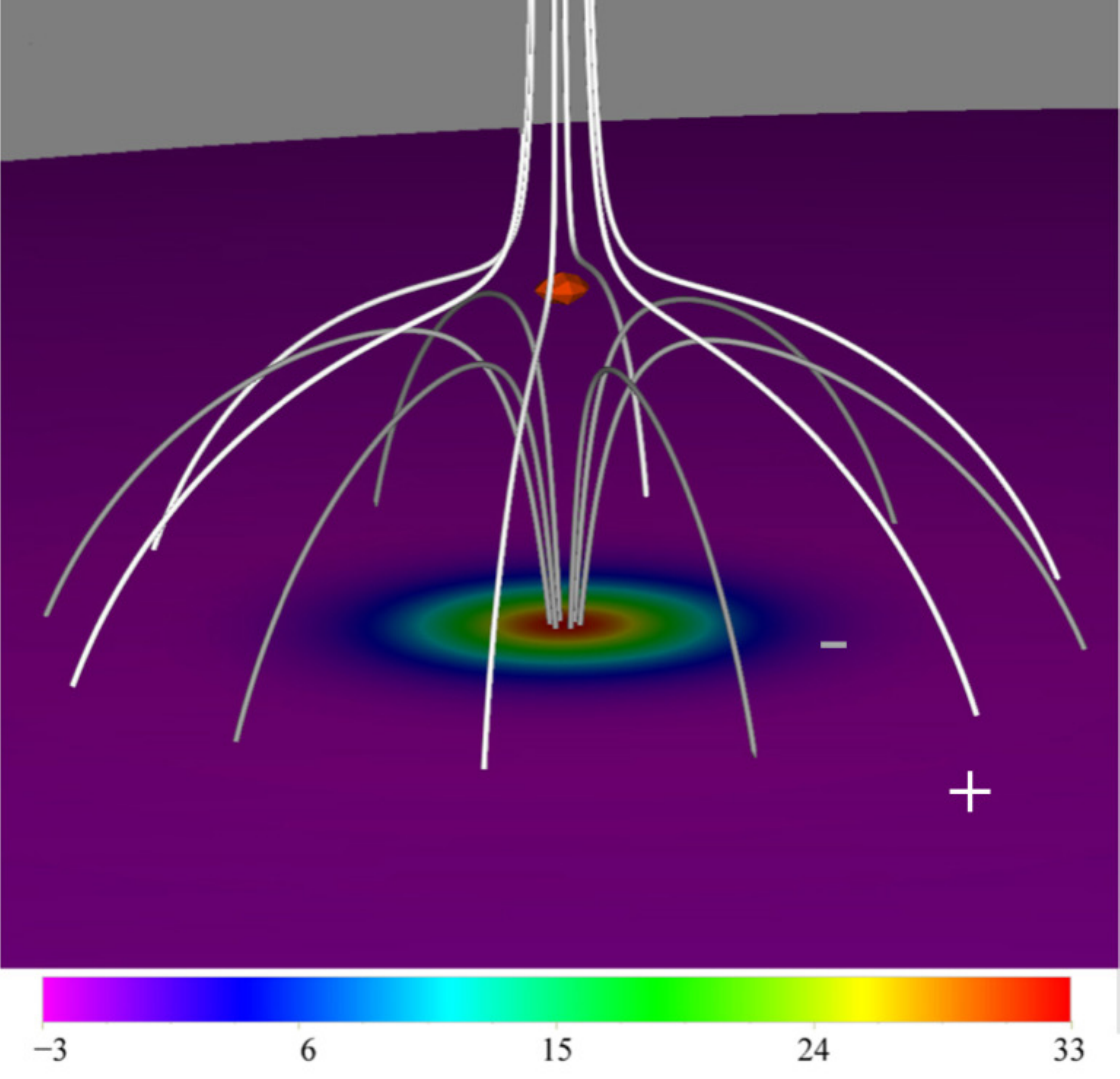}
\caption{A view of the embedded minority-polarity concentration and the surrounding majority-polarity corona at $t=0$.  Color scale denotes the magnitude and sign of the radial magnetic field at the solar surface in Gauss.  White and gray curves respectively show magnetic field lines in open and closed flux systems, indicating the location of the separatrix surface, with the plus and minus symbols marking the opposing polarities. The high-$\beta$ region $(\beta \equiv P_{\rm{thermal}} / P_{\rm{magnetic}})$ encompassing the magnetic null point of the system is shown as a closed red surface suspended between the two flux systems.}
\label{fig:jet_driver}
\end{figure}

The coronal-hole jet simulation that we performed and analyzed for this study is a straightforward extension of the K17 calculation, whose results were analyzed further by U17.  For brevity, we summarize the main features of the simulation below, while we refer the reader to K17 for more details.
The full simulation domain in spherical coordinates $(r,\theta,\phi)$ is $[1R_\sun,60R_\sun]\times[-9^\circ,+9^\circ]\times[-9^\circ,+9^\circ]$ in radial coordinate $r$, latitude $\theta$, and longitude $\phi$.  This is the same angular extent as K17, but with the outer radial boundary moved from 9$R_\sun$ to 60$R_\sun$.  The grid is uniformly spaced in the angular directions $\theta$ and $\phi$ and 

in $\ln r$ in the radial direction $r$.  The inner and outer radial boundaries of the domain are open, while the four side boundaries are closed.

The time-dependent, ideal, compressible, isothermal, MHD equations for mass, momentum, and magnetic field were solved using the Adaptively Refined Magnetohydrodynamics Solver \citep[ARMS;][]{devore2008}.  The PARAMESH toolkit \citep{macneice2000,olson2005} managed the parallel computations and the adaptive grid refinement process, which targeted regions of relatively strong electric current density for better resolution.   The plasma was assumed to be isothermal ($T_\sun = 10^6$ K), eliminating the MHD energy equation and providing a straightforward, efficient way to produce a background supersonic solar wind \citep{parker1958}.  The temperature ($T_\odot$), as well as the remaining model parameters, such as the base mass density ($\rho_\odot = 2.0 \times 10^{-13}$ kg m$^{-3}$), were chosen to reflect routinely observed environmental conditions for solar CH jets.

The magnetic configuration of the jet-driving region is shown in Figure \ref{fig:jet_driver}.  A radially oriented background field, uniform in the transverse dimensions, was assumed ($B_b = -2.5$ G at 1$R_\sun$).  A radially oriented magnetic dipole of opposite polarity (maximum $B_d = +35$ G at 1$R_\sun$) was placed at a depth $R_\sun / 70$ below the solar surface.  These superposed fields form a dome-shaped separatrix surface with a null point at the top of the closed-flux region, as shown in the figure, and a circular polarity inversion line (PIL) at the solar surface within the dome.  A slow rotational motion about the center of the embedded bipole was imposed inside the PIL in order to twist and energize the magnetic field under the separatrix.  Eventually, the accumulated twist exceeded the threshold for kink instability, causing the field beneath the dome to topple and reconnect rapidly through the null point above the dome, driving the resultant CH jet along adjacent magnetic field lines.  The imposed surface motion was ramped up from rest during the first 1000s, then held fixed at peak speed for 1500s, and finally ramped back down to rest for a further 1000s.  Thereafter, the plasma and magnetic field lines were held fixed at the surface.  The magnetic energy released and the kinetic energy generated by the jet were essentially unchanged from the values reported by K17.  Nonlinear Alfv\'{e}n waves and plasma flows generated by the violent kink-driven reconnection comprised the jet, which propagated into the outer corona and inner heliosphere. The evolution was the same as for the K17 jet: here we tracked it over a much longer time interval (21000 s vs.\ 4000 s in K17), during which the Alfv\'{e}n wave front propagated much further (38$R_\sun$ vs.\ 5$R_\sun$ in K17), as did the leading edge of the trailing dense jet (13$R_\sun$ vs.\ 1.5$R_\sun$).

\section{Parker Solar Probe}
\label{sec:psp}

\begin{figure}
\includegraphics[width=\linewidth]{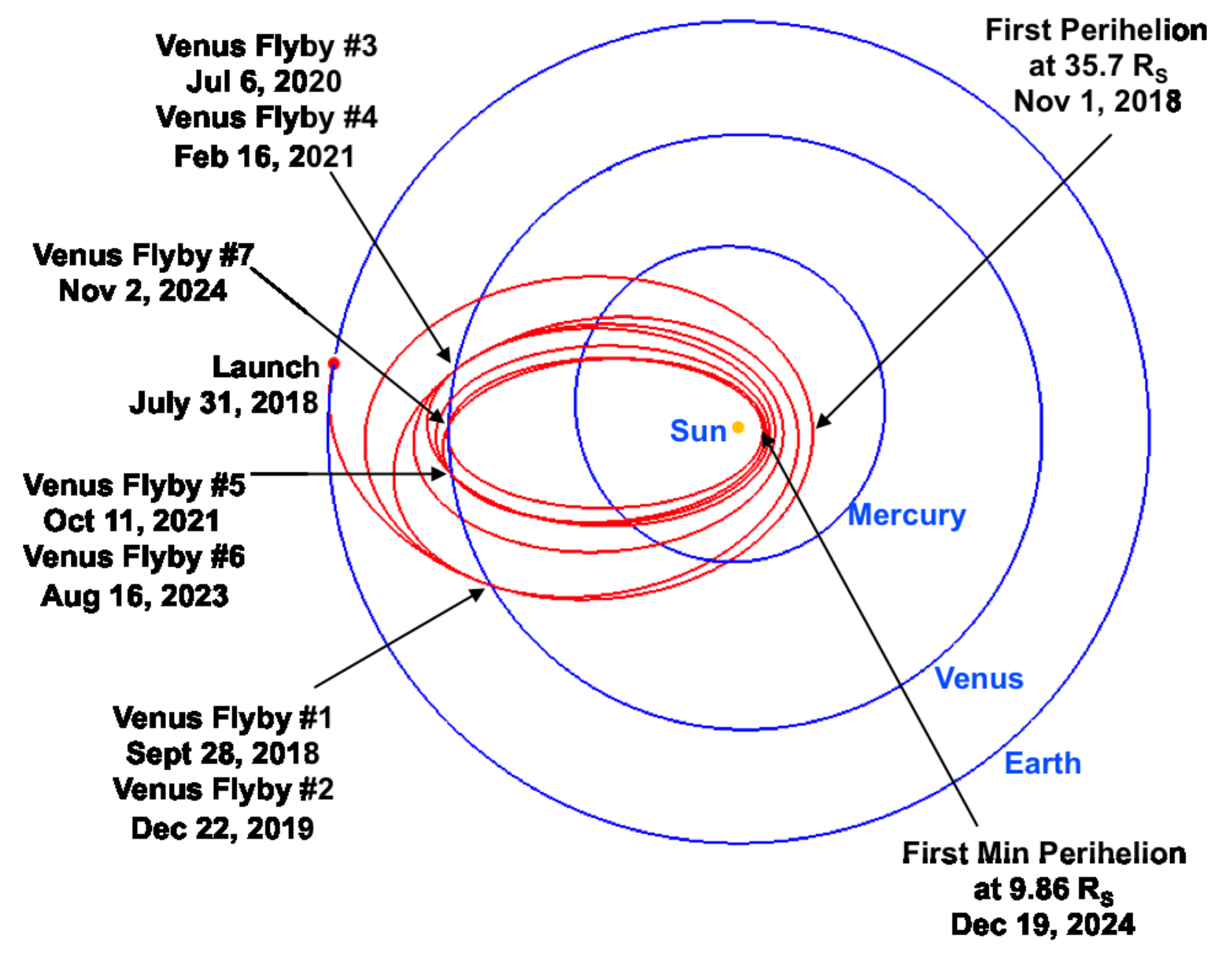}
\caption{Illustration of PSP's planned trajectory from \citet{fox16}.  The red line traces the orbital path, and the blue circles show the orbits of Mercury, Venus, and Earth. The elliptical PSP orbits become more eccentric and reach progressively lower perihelia due to seven planned Venus gravity assists over the course of the mission.}
\label{fig:psp_orbit_plot}
\end{figure}
\begin{figure*}
\begin{center}
\includegraphics[width=15cm]{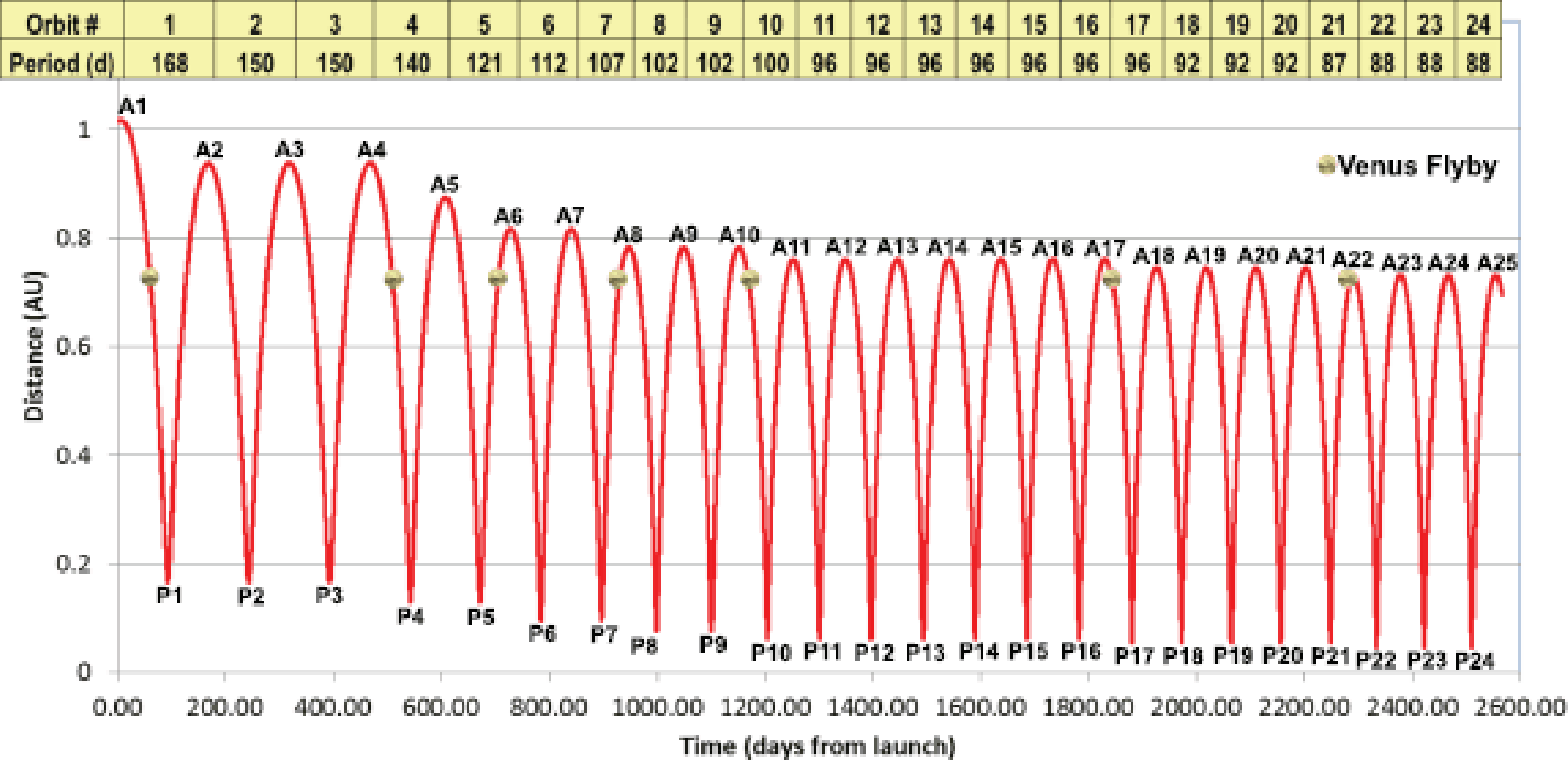}
\caption{Distance vs.\ time for the PSP mission, from \citet{fox16}.  Each column gives an orbit number and period aligned with that orbit's portion of the time/distance plot.  A small image of Venus is placed at each point where PSP executes a gravity-assist maneuver.  The aphelion and perihelion for each orbit are marked with `A' and `P,' respectively, and the orbit number.}
\label{fig:psp_orbit_vs_time}
\end{center}
\end{figure*}

PSP is scheduled to launch in July 2018 and become the first spacecraft to provide \textit{in situ} measurements of the inner heliosphere within 10$R_\sun$.  It will take up an orbit nearly in the ecliptic plane and, with the help of seven Venus gravity assists, will reduce its perihelion altitude from initially 35$R_\sun$ to eventually 8.86$R_\sun$ over the course of the seven-year nominal mission. PSP is expected to spend a total of 937 hours inside 20$R_\sun$ and 14 hours inside 10$R_\sun$.  The close approach and extended duration below 20$R_\sun$ will allow PSP to measure the coronal conditions that drive the nascent solar wind and detect eruptive transients such as CH jets.  The mission will carry four instrument science investigations: the Electromagnetic Fields Investigation \citep[FIELDS;][]{bale16} to measure electromagnetic fields and plasma-density fluctuations; the Integrated Science Investigation of the Sun \citep[ISIS;][]{mccomas16} to observe accelerated particles at energies from tens of keV to 100 MeV; the Solar Wind Electrons Alphas and Protons Investigation \citep[SWEAP;][]{kasper16} to obtain particle counts for electrons, protons, and helium ions; and the Wide-field Imager for Solar Probe Plus \citep[WISPR;][]{vourlidas16} white-light telescope to image the corona and inner heliosphere.  Figure \ref{fig:psp_orbit_plot} shows the orbital paths throughout the mission. The effects of the gravity assists on the orbital period, aphelion, and perihelion are displayed in Figure \ref{fig:psp_orbit_vs_time}. Our study focuses on detections of density, velocity, and magnetic field during the final three orbits (22-24) in the mission, all of which occur after the final gravity assist and attain perihelia inside 10$R_\sun$. These orbits begin in late 2024 and provide the best chance for PSP to encounter equatorial coronal holes. The Sun then will be approaching solar maximum, when equatorial coronal holes and the CH jets they contain are observed most frequently.

\section{Flying Through the Model}
\label{sec:ftm}
\subsection{Grid Regularization}
\label{sec:grid_reg}
The adaptive grid utilized by ARMS is essential to resolve adequately important jet features in the model, but it presents formidable challenges to a quantitative analysis of those features.  We found that a regularized grid was necessary in order to establish consistent spacecraft coordinates and properly analyze the physical properties of the simulated CH jet.  The algorithm chosen had to regularize the grid while maintaining the accuracy of the irregularly gridded data. For its simplicity and strong weighting of nearest-neighbor points in determining the new grid values, we chose linear interpolation based on a \citet{delaunay34} triangulation of the irregular points. This was implemented in the Interactive Data Language (IDL) using the \textit{Triangulate} and \textit{Trigrid} functions, as follows.

\begin{figure*}
\includegraphics[width=18cm]{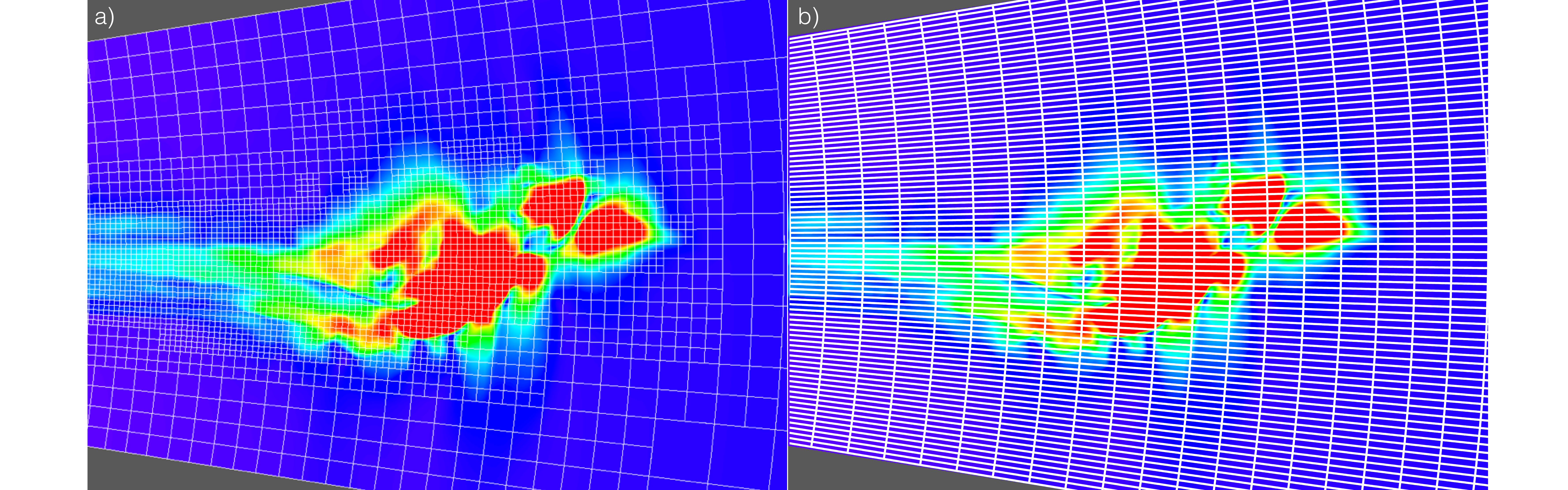}
\caption{An illustration of the grid regularization procedure described in \S\ref{sec:grid_reg}. The total velocity ($|V|$) in the $\phi=0$ plane is shown at $r\approx$6-10 $R_{\odot}$ and $t$ = 6125 s in both images. (a) The ARMS adaptive grid. (b) The grid resulting from our regularization process. In both panels only every fourth grid line is rendered, for clarity.}
\label{fig:grid_reg}
\end{figure*}

After selecting a given two-dimensional slice of the model space at constant radius $r$, the irregularly gridded $\theta$ and $\phi$ coordinates were read into IDL.  From this set of co-planar points, the \textit{Triangulate} procedure constructs the Delaunay triangles using the `divide and conquer' method \citep{lee80}. The procedure generates a set of triangles from the given set of irregular points, such that a circle formed by connecting the vertices of any triangle does not contain any other point.  This maximizes the minimum interior angle of the triangles and ensures that only nearby points are used to construct them. After they have been created, a spatially regular grid is overlaid on the triangles by the \textit{Trigrid} function. The value at each point of this regular grid is given by a linear interpolation of the values at the vertices of the Delaunay triangle into which it falls. This ensures that the values used in the interpolation are those closest to the specified regular grid point, thereby minimizing the distortion of the original dataset.  The complete procedure was performed at regularly spaced values of $r$ ($\Delta r = 0.05R_{\odot}$ = 35000 km) to build a uniformly gridded dataset spanning the range of values required for a given fly-through trajectory (see Figure \ref{fig:grid_reg}).

The sampling time for our analysis was chosen in accordance with the coarse-grained resolution of the regularized grid based on the following considerations.  All jet features of interest except short-lived transients at the shock front had characteristic velocities below $V_{max} \approx 1400$ km s$^{-1}$. The most demanding time-cadence requirement is associated with the smallest features occupying a single grid cell. Assuming that a feature has a spatial extent of $\Delta r$ (one grid cell in $r$), the sampling time $\Delta t$ needed to resolve such a feature as it moves radially outward with a speed of 1400 km s$^{-1}$ is $\Delta t \approx \Delta r / V_{max}$ = 25 s. Hence, the sampling time for this study was chosen to be $\Delta t$ = 25 s. 

\begin{figure}
\includegraphics[width=\linewidth]{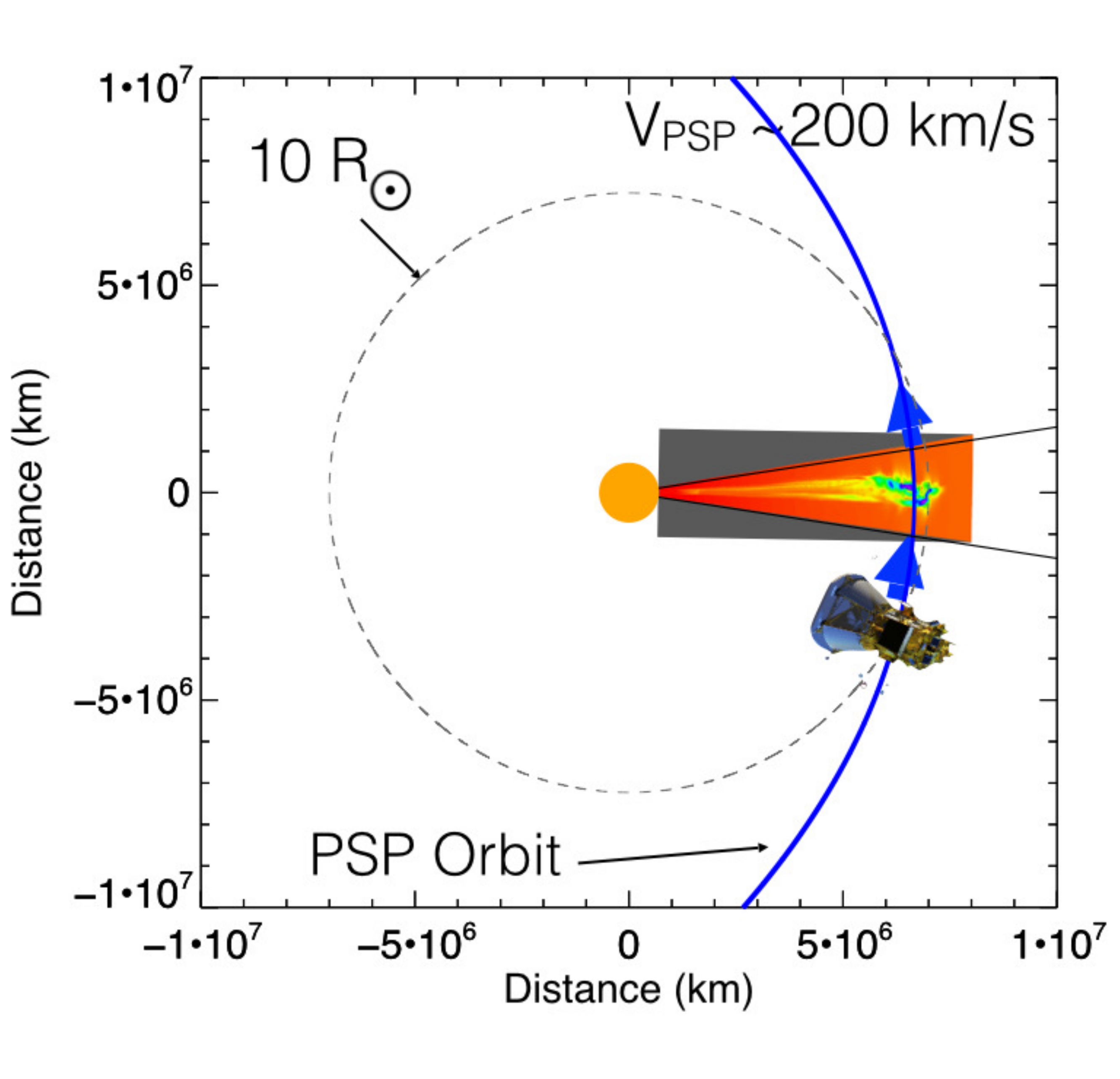}
\caption{A cross-section (at $\phi=0$) of the model space (red-orange wedge) superimposed over the fitted path of PSP during orbits 22-24 (blue ellipse). An image of PSP (not to scale) is shown moving in the direction of the blue arrows.  The Sun (to scale) is shown as an orange circle.}
\label{fig:psp_orbit_fit}
\end{figure}

\subsection{Interpolated Orbital Trajectory}
\label{subsec:orbits}
The trajectories that we use are derived from preliminary projected orbital parameters developed by the PSP mission team (A.\ Szabo, personal communication).  The dataset gives the position of PSP once per day throughout the mission in Heliocentric Mean Ecliptic J2000 coordinates.  We converted to spherical coordinates and rotated the domain so that the PSP orbital plane coincides with the $\phi=0^\circ$ plane in our model space. Because the final three orbits (orbits 22-24) are nearly identical and the data points are sparse, especially near perihelion where the spacecraft moves most rapidly, data points from all three orbits were used to find a best-fit elliptical path, shown in Figure \ref{fig:psp_orbit_fit}.  We then restricted the path to the 18$^\circ$ angular size of our model space and interpolated data points along the path near perihelion corresponding to a data measurement cadence of 1 s at a spacecraft speed of 190 km s$^{-1}$, to emulate the conditions expected for PSP.  The result was an array of spacecraft locations in spherical coordinates at 1-s intervals along the fitted elliptical orbit.

In the less than 9$^\circ$ wide segment of PSP's projected orbital path crossing the jet, the radial position $r$ satisfies $9.516R_\sun \le r \le 9.541R_\sun$. The most recent estimates for the PSP mission now project a minimum perihelion at 8.86$R_\sun$, slightly lower than the assumed value for this simulation.  Given the persistent structure of our CH jet and the relatively small change in perihelion, this modification does not significantly affect the simulation results presented below.

Taking the PSP orbital trajectory into account is necessary because the characteristic speeds of the PSP spacecraft and the CH jet flows are comparable.  Both speeds are on the order of a few hundred km s$^{-1}$, so the encounter cannot be captured well using either a simple fixed-$t$ spatial path through the model space (assuming $V_{PSP} \gg V_{Jet}$) or a fixed-{\bf r} time history (assuming $V_{PSP} \ll V_{Jet}$).  Therefore, we adopted an approach that accounts for the motions of both the spacecraft and the jet in the following analysis.

\subsection{Fly-Through Calculations}
\label{subsec:ftc}
To simulate likely PSP jet encounters, four prototypical trajectories were studied: one passing through each of the three separate jet regions identified by U17 (see \S\ref{sec:results} and Figure \ref{fig:jet_cartoon_ft} below), and one co-rotational trajectory that sampled the entire length of the jet.  The first three encounter scenarios have PSP flying along the elliptical arc through the model space described above, differing only in its start time during the simulation to sample the various regions.  The simulated spacecraft followed its orbital path across the jet in each case.  For each four-dimensional spacetime position, the algorithm reads the time stamp of the spacecraft and opens the closest available time step from the model output.  The radial coordinate of PSP is then read, and the appropriate surface of constant $r$ is retrieved from the regularly gridded output.  Next, the $\theta$ and $\phi$ trajectory coordinates are used to determine which model pixel currently contains the spacecraft.  Finally, the data values at that pixel location are returned.  This process is applied repeatedly to generate a time history for each variable along the PSP trajectory. Although the orbital path has a temporal cadence of 1s, PSP moves at only $\approx $200 km s$^{-1}$. Hence, at the 25s temporal cadence of the jet model output, PSP moves about 5000 km, much less than the 35000-km spacing of the regularized grid.  Consequently, the virtual spacecraft samples the same pixel multiple times, even though it is temporally progressing through the model space.  In order to smooth the transitions in the output as PSP moves from pixel to pixel, we applied a 15s wide boxcar average to each of the plotted parameters.

\begin{figure}
\begin{center}
\includegraphics[width=\linewidth]{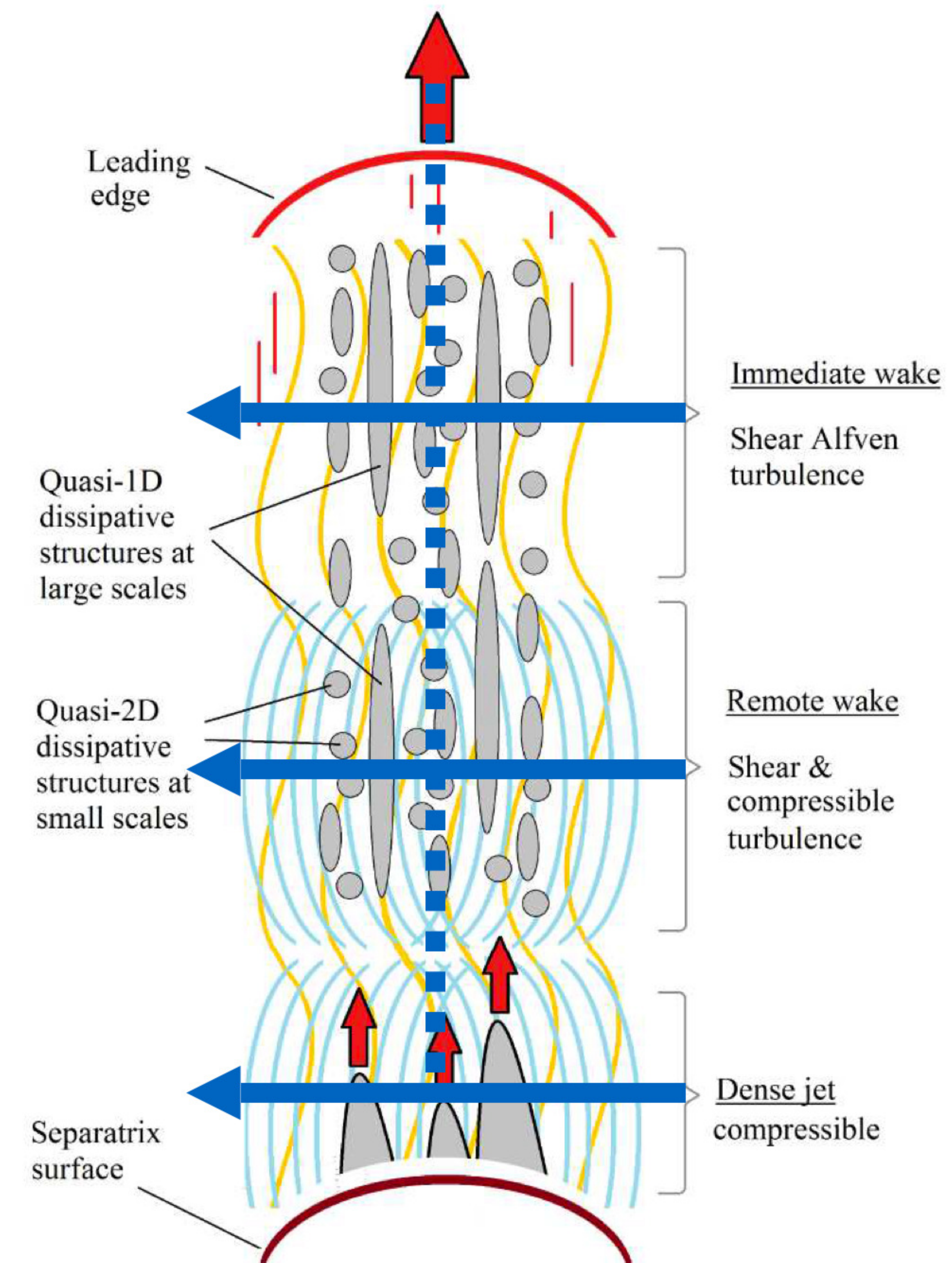}
\caption{Schematic diagram showing the three regions of distinctive internal structure from the analysis of \cite{uritsky17}.  Red arrows mark the direction of plasma flow; blue arrows indicate schematically the three simulated PSP encounters through those regions.  The dotted blue line illustrates the path of the co-rotational encounter and approximates the jet axis.  Regions of Alfv\'{e}nic and compressible wave activity are marked by yellow and light blue curves, respectively.}
\label{fig:jet_cartoon_ft}
\end{center}
\end{figure}

Although our four prototypical trajectories represent idealized cases that are unlikely to be encountered in practice, they span the fundamental space of all encounters that PSP is expected to have with a CH jet. For example, any perpendicular fly-through that misses the jet axis would sample only part of the envelope, while the core would be sampled more briefly or not at all. Obviously, this would abbreviate the duration of measurements of the conditions inside the jet. On the other hand, a sufficiently oblique fly-through by PSP would prolong the duration of measurements beyond any of our perpendicular fly-throughs; the most extreme example of this is the co-rotational encounter presented below. A different oblique fly-through might sample only two, or just one, of the three distinct regions that characterize our jet.

The schematic Figure \ref{fig:jet_cartoon_ft} is adapted from U17, with the simulated PSP spacecraft trajectories shown as three solid blue arrows and the dashed blue line.  The arrow trajectories sample the jet's immediate wake (top arrow, earliest time), remote wake (middle arrow, intermediate time), and dense jet (bottom arrow, latest time) regions separately.  The dashed-line trajectory is for a Co-rotational fly-through, in which the spacecraft hovers over a fixed location on the Sun and the jet passes over PSP, so the jet is sampled from the leading edge progressively through the other three regions, top to bottom.  These four cases are discussed in turn.

\section{Results}
\label{sec:results}

\subsection{Case 1: Immediate Wake}
\label{subsec:imwake}

\begin{figure*}
\begin{center}
\includegraphics[width=14.5cm]{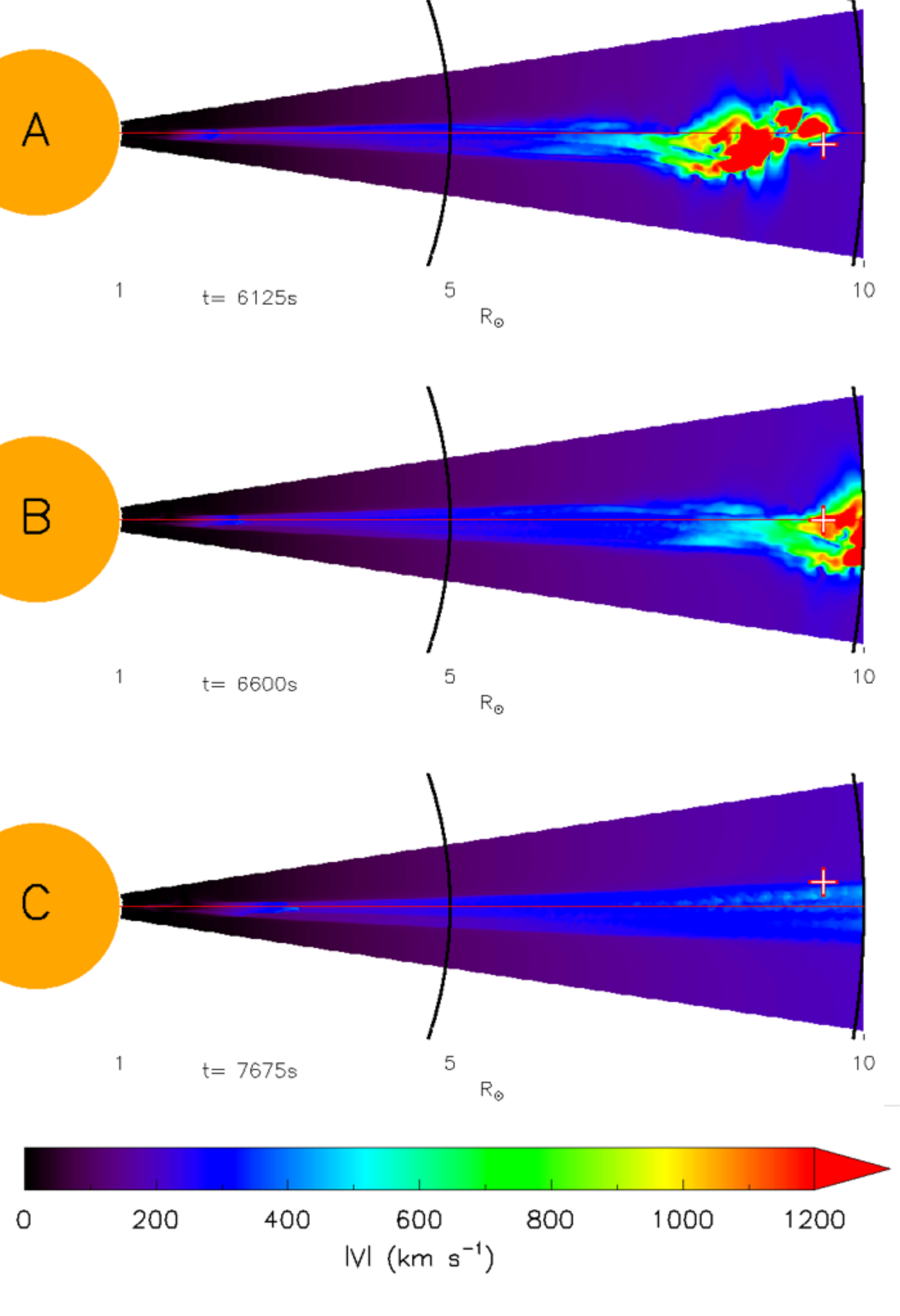}
\caption{Case 1: Images of total velocity ($|V|$) in the $\phi=0$ plane at selected times during the simulation, illustrating a PSP encounter with a CH jet during the immediate wake fly-through. The red and white crosses mark the instantaneous PSP position. The images span $r \in \left[1R_\sun,10R_\sun\right]$ and $\theta \in \left[-9^\circ,+9^\circ\right]$. Each image (A,B,C) corresponds in time to the identically labeled dashed vertical line drawn in Figure \ref{fig:implots}.}
\label{fig:imimages}
\end{center}
\end{figure*}

\begin{figure*}
\begin{center}
\includegraphics[width=17cm]{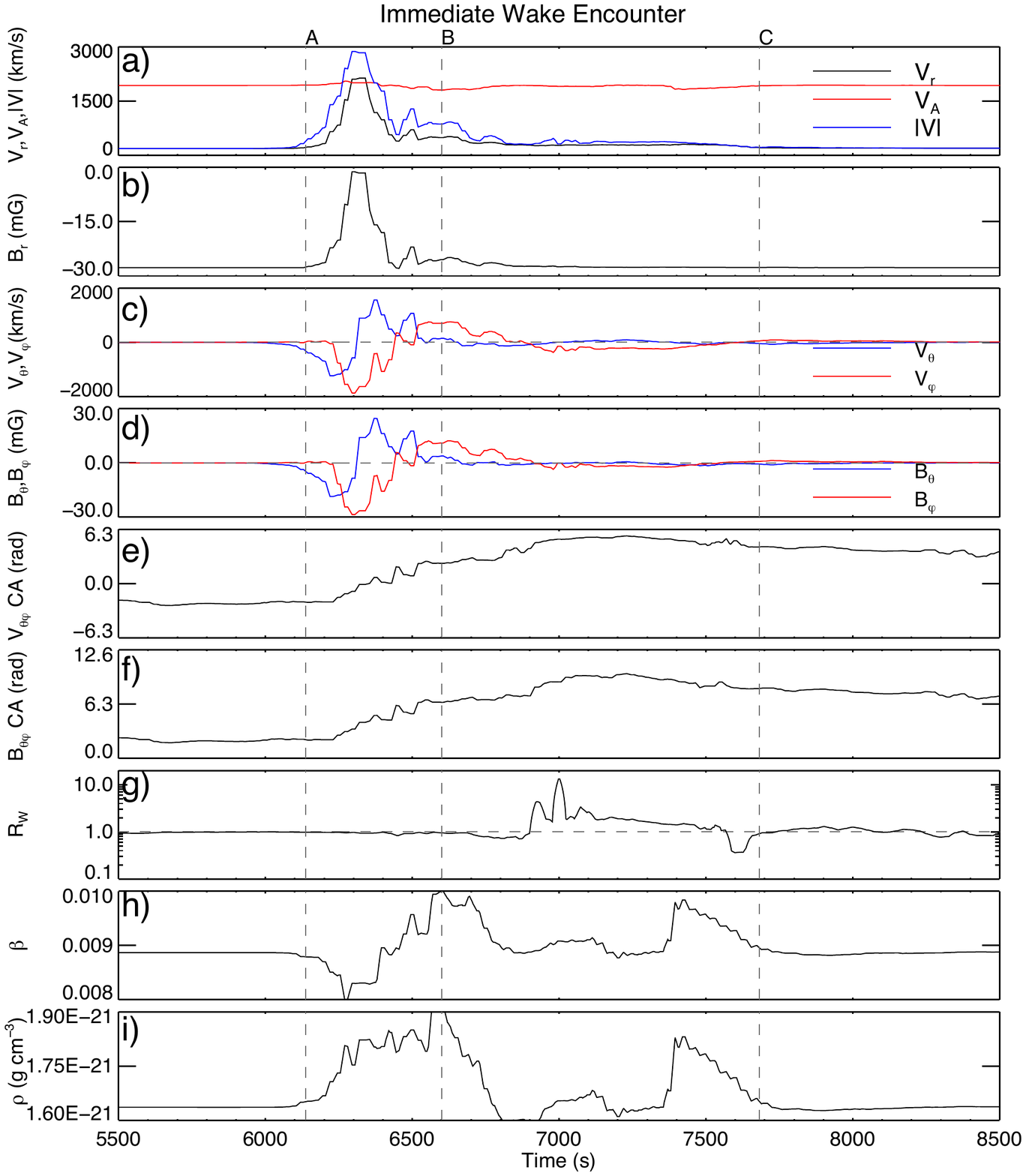}
\caption{Time series of simulated PSP encounter with the CH jet immediate wake.  From top to bottom: a) radial velocity V$_r$, Alfv\'{e}n speed V$_A$,and total velocity $|V|$, b) radial magnetic field B$_r$, c) tranverse velocity components V$_\theta$ and V$_\phi$, d) transverse magnetic field components B$_\theta$ and B$_\phi$, e) clock angle of the transverse velocity vector V$_{\theta\phi}$, f) clock angle of the transverse magnetic field vector B$_{\theta\phi}$, g) Wal\'{e}n ratio R$_w$, h) ratio of gas to magnetic pressure $\beta$, and i) mass density $\rho$. Each dashed vertical line indicates a time at which an image is presented in Figure \ref{fig:imimages}. Figures \ref{fig:rmplots}, \ref{fig:djplots}, \ref{fig:corotplots}, \ref{fig:offaxisplots} follow the same format.}
\label{fig:implots}
\end{center}
\end{figure*}

In this simulated encounter, the spacecraft samples the region directly behind the nonlinear Alfv\'{e}n wave front that leads the jet (top blue arrow, Fig.\ \ref{fig:jet_cartoon_ft}).  This is U17's immediate wake.  Figure \ref{fig:imimages} images the model space for $r \in \left[1R_\sun,10R_\sun\right]$ and $\theta \in \left[-9^\circ,+9^\circ\right]$ at $\phi = 0^\circ$.  Color shading renders the velocity magnitude, $|V|$.  The simulated PSP spacecraft begins at coordinates $\left[9.54R_\sun,-4.5^{\circ},0^{\circ}\right]$ at time $t=3900$ s, and follows the trajectory described above.  The spacecraft first encounters the jet at time $t=6137$ s, soon after Figure \ref{fig:imimages}A, and enters the region of highest flow speed at $t\approx6300$ s.  Subsequently, the spacecraft moves into a striated region of lower speeds, in which local concentrations of enhanced velocity trail behind the leading front.  PSP exits the jet at $t=7681$ s, just after Figure \ref{fig:imimages}C.  Defining the effective width $w_{eff}$ as the distance traversed by the spacecraft inside the jet, we find that the spacecraft spends $1544$ s inside the jet, yielding $w_{eff} \approx 2.9 \times 10^5$ km = 0.42$R_\sun$.

The simulated PSP \textit{in situ} measurements for this encounter are shown in Figure \ref{fig:implots}.  At $t\approx6100$ s, the spacecraft encounters the jet boundary, defined as where the radial velocity $V_r$ is enhanced at least 20 km s$^{-1}$ above the average steady background velocity of the radially flowing ambient solar wind.  At the spacecraft altitude, the average background $V_r$ is 192 km s$^{-1}$.  As the spacecraft enters the jet, it passes into a region of enhanced $V_{r}$, rapidly increasing to more than an order of magnitude above the background solar wind radial velocity to a peak value of 2135 km $s^{-1}$.  $V_r$ seems to exceed the local Alfv\'{e}n speed $V_A$ (shown by the red line in Figure \ref{fig:implots}a) inside the leading edge of the jet.  However, when the motion is translated into the rest frame of the ambient solar wind, which is moving at a speed 192 km s$^{-1}$, the Alfv\'{e}n Mach number $M_A={V_r}/{V_A} \approx 0.97$. This indicates the presence of an Alfv\'{e}nic shock front caused by the nonlinear, torsional Alfv\'{e}n wave released at jet initiation, as seen also in the K17 jet, and moving at approximately the expected local Alfv\'{e}n speed. In the region where $V_r$ increases to the local Alfv\'{e}n speed, the local magnetic field vector is strongly deflected from the radial direction by the nonlinear wave (Fig.\ \ref{fig:implots}b).  As a result the radial magnetic field approaches zero in this region (Fig.\ \ref{fig:implots}b), in contrast to the rest of the simulation space both inside and outside the jet, where the radial component dominates. The change in the radial magnetic field $B_r$ (Fig.\ \ref{fig:implots}b) is closely correlated with the radial velocity.  While $V_A$ apparently remains constant throughout the traversed region when compared to the jet velocity, it does in fact experience small variations due to the \textit{B}-field fluctuations. These fluctuations are on a much smaller scale than that of the radial velocity, since the changes in \textit{B} and $\rho^{1/2}$ are small and positively correlated, with the result that their ratio stays nearly the same.

This shock is also accompanied by the onset of perturbations in the velocity and magnetic field components transverse to the jet outflow (Fig.\ \ref{fig:implots}c,d), 
\begin{equation}
V_{\theta\phi} \equiv (V_\theta^2 + V_\phi^2)^{1/2},
B_{\theta\phi} \equiv (B_\theta^2 + B_\phi^2)^{1/2}.
\end{equation}
These transverse perturbations are further reflected in the clock angle (CA) of  $V_{\theta\phi}$ and $B_{\theta\phi}$ vectors (Fig.\ \ref{fig:implots}e,f).  Here the clock angle remains steady until the encounter with the jet front, where it undergoes more than a full rotation from $t=6200$ s to $t=7200$ s.  $V_{\theta\phi}$ and $B_{\theta\phi}$ are negligible upstream from the shock front, and suddenly take on large, non-zero values downstream from the shock.  In addition, the magnetic pressure increases across the shock, indicating a MHD fast-mode switch-on shock front \citep[see, e.g.,][]{kulsrud05}, as deduced previously for the K17 jet \citep{devore2016}.  These large values of $V_{\theta\phi}$ combine with the enhanced $V_r$ to produce a maximum $|V|$ of $2872$ km s$^{-1}$ in the shock region.

The \citet{walen1944} number is the ratio of the transverse plasma flow speed to the Alfv\'{e}n speed associated with the transverse magnetic field, 
\begin{equation}
R_w\equiv(\mu_0\rho)^{1/2}\Delta V_\perp/\Delta B_\perp.
\end{equation}
For an ideal linear or nonlinear Alfv\'{e}n wave in an isotropic plasma, $R_w = \pm 1$.  As shown in Figure \ref{fig:implots}g, $R_w \approx 1$ throughout the encounter with the jet shock front and its immediate wake region.  It deviates from this value beginning around $t=6900$ s, when PSP emerges from the region of strongly enhanced $V_{\theta\phi}$ and $B_{\theta\phi}$.  The average $R_w$ in this region is 0.955, closely fulfilling the condition for a purely Alfv\'{e}nic region in an isotropic plasma. 

The spacecraft remains in a region of greatly increased $V_r$ from $t \approx 6100$ s to $t \approx 6900$ s. There are two successive enhancements in mass density and plasma $\beta$ associated with strong excursions in $R_w$, one well above and one well below unity. 
Thereafter, the virtual spacecraft remains in a region of slightly increased $V_r$ until $t \approx 7700$ s, when the values of both the velocity and magnetic field return to levels comparable to those observed before the jet encounter.  These signatures suggest an exterior sheath or envelope of mixed jet and ambient plasma, with small but finite transverse velocity and magnetic fields surrounding the jet.  The virtual spacecraft spends about $900$ s inside the strongly jetting plasma itself and $2200$ s inside the surrounding jet envelope.

We note that all three components of the velocity are highly correlated with the corresponding components of the magnetic field (Fig.\ \ref{fig:implots}a-d).  This correlation also reinforces the conclusion that the immediate wake is a region of incompressible, purely Alfv\'{e}nic flow.  The density is enhanced by 18\% over the background when the spacecraft passes into the jet region, accompanied by a dip in the plasma $\beta$ (Fig.\ \ref{fig:implots}h,i), but this front does not carry the bulk of the jet's dense plasma with it. This ejected dense plasma is left behind by the Alfv\'{e}nic wave front, as discussed in \S\ref{subsec:djregion}.

To summarize, in the simulated encounter with the immediate wake, we find a sudden, simultaneous increase in all three velocity components as the spacecraft encountered this jet region.  There is no concurrent large increase in $\rho$, indicating that the wavefront is weakly compressible.  A decrease in the radial component of the magnetic field is observed simultaneous with and well correlated to the change in $V_r$.   The onset of fluctuations in the transverse velocity and magnetic field components concurrent with the decrease in $B_r$ and a steady value of $\lvert$B$\rvert$ in this region is consistent with the angle of the magnetic field changing across this feature, suggesting that this wavefront is a weak switch-on type shock.  Further, $V_{\theta\phi}$ and  $B_{\theta\phi}$ show full rotations in their clock angles, revealing the helical interior structure of the jetting plasma.  The Wal\'en number remaining at a value of $+1$ confirms that the plasma flow inside the immediate wake is Alfv\'{e}nic and therefore incompressible.

\subsection{Case 2: Remote Wake}
\label{subsec:rmwake}

\begin{figure*}
\begin{center}
\includegraphics[width=14.5cm]{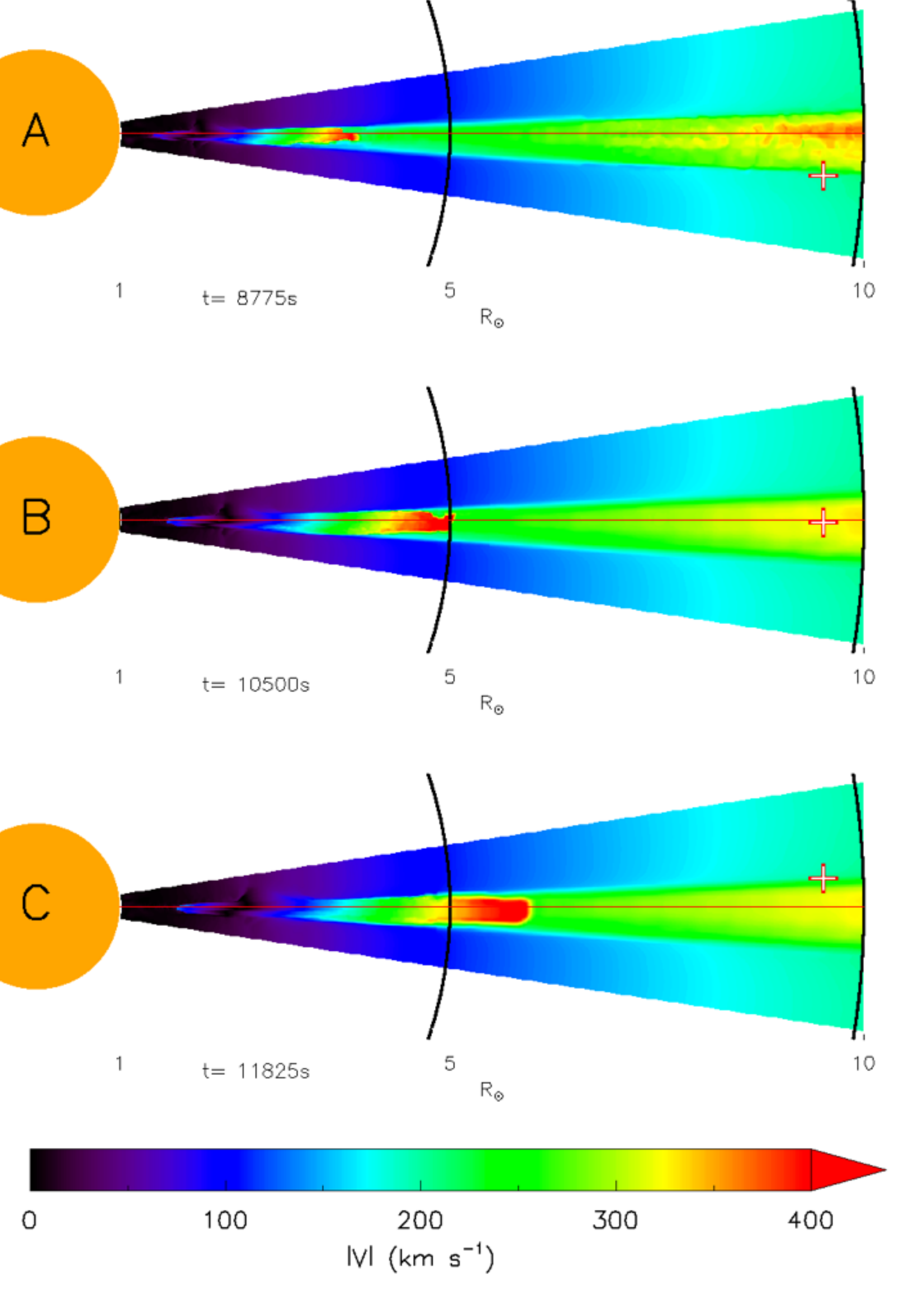}
\caption{Case 2: Same as Figure \ref{fig:imimages} for the remote wake fly-through. Each image (A, B, C) corresponds in time to the identically labeled dashed vertical line drawn in Figure \ref{fig:rmplots}.}
\label{fig:rmimages}
\end{center}
\end{figure*}

\begin{figure*}
\begin{center}
\includegraphics[width=17cm]{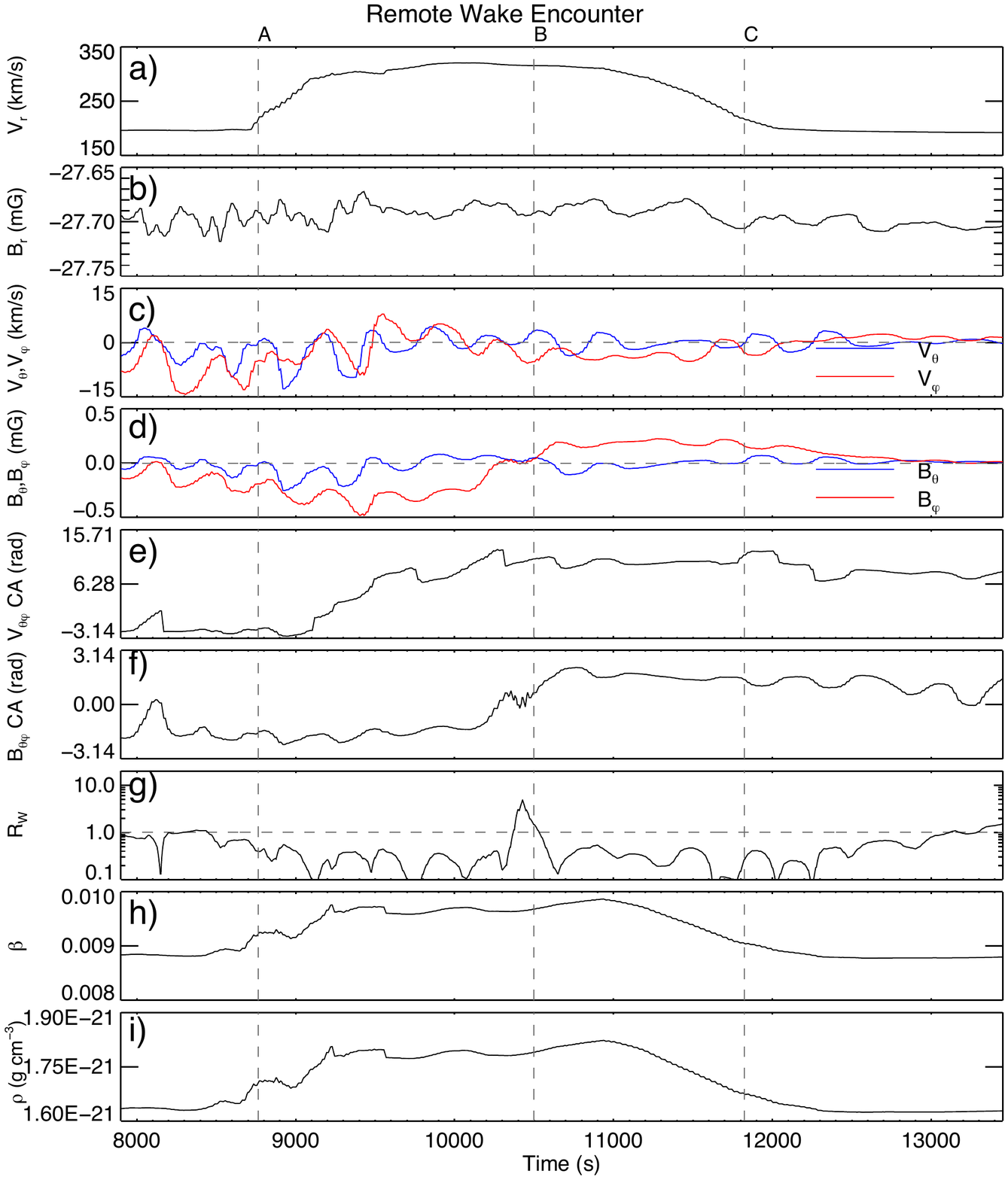}
\caption{Time series of plasma variables for the remote wake fly-through.  Each dashed vertical line indicates a time at which an image is presented in Figure \ref{fig:rmimages}.}
\label{fig:rmplots}
\end{center}
\end{figure*}

The remote wake of the jet is a mixed region, as described in U17, which is less dynamic compared to the shock front and the immediate wake.  Figure \ref{fig:rmimages} shows the spacecraft passing into and out of a jet region that, while still distinguishable from the background solar wind, is significantly more homogeneous than the leading portion of the jet.  The simulated PSP measurements are shown in Figure \ref{fig:rmplots}. The simulated PSP spacecraft begins at coordinates $\left[9.54R_\sun,-4.5^{\circ},0^{\circ}\right]$ at time $t=7900$ s, and follows the calculated trajectory.  PSP encounters the jet at $t=8764$ s, and exits the jet at $t=11822$ s, spending $3058$ s inside the structure; the resulting jet width is $w_{eff} \approx 5.8 \times 10^5$ km = 0.83$R_\sun$.  $V_r$ increases from a background value of 194 km s$^{-1}$ to a maximum of 320 km s$^{-1}$ within the jet.  The 126 km s$^{-1}$ enhancement in $V_r$ (Fig.\ \ref{fig:rmplots}a) compared to the background is measurable and significant, clearly indicating the boundaries of the jet. The $B_{r}$ component (Fig.\ \ref{fig:rmplots}b) maintains a steady value within 0.1\% throughout the encounter, unlike the immediate wake in which the magnetic field vector was strongly deflected from its ambient (radial) direction.

In this case, the fluctuations in $V_{\theta\phi}$ and $B_{\theta\phi}$ (Fig.\ \ref{fig:rmplots}c,d) are small and significantly less correlated than in the immediate wake region.  There is a small-scale, quasi-periodic oscillation (with a period of about 330s) which becomes less coherent as the encounter progresses in time. There is also a large-scale oscillation, most apparent in the $B_{\phi}$ component, which indicates that $B_{\phi}$ varies spatially inside the jet, differing in sign on either side of the jet's approximate center.  The $B_{r}$ component maintains a steady value throughout the encounter, in contrast to the previous case.  The $V_{\theta\phi}$ and $B_{\theta\phi}$ clock angles (Fig.\ \ref{fig:rmplots}e,f) have more and larger oscillations than in the previous case.  Both still show a large-scale rotation across the jet, but the rotation in $V_{\theta\phi}$ significantly precedes that in $B_{\theta\phi}$.  The Wal\'en number $R_w$ in the remote wake is initially close to one, but deviates substantially from unity as the encounter with the jet continues (Fig.\ \ref{fig:rmplots}g), indicating that unlike the immediate wake, the remote wake is not a region containing a purely incompressible, Alfv\'{e}nic flow.  The plasma $\beta$ (Fig.\ \ref{fig:rmplots}h) does not decrease as it did in the encounter with the initial shock front; instead it rises and falls, reflecting the mild density changes. Comparable to the immediate wake, the density $\rho$ (Fig.\ \ref{fig:rmplots}i) increases only 12\% in this region, indicating that the plasma does not experience much compression here, either.

The correlation between the respective components of the velocity and magnetic field seen in the immediate wake is notably absent here, with the exception of the small-scale oscillations in the transverse components mentioned earlier, whose perturbations correlate quite well, especially in the early part of the encounter.  These findings reinforce our conclusion (U17) that the remote wake is a mixed region that shares characteristics of both the Alfv\'{e}n wave front and immediate wake that precede it and the dense jet region that follows it. Our simulations also demonstrate that these mixed properties persist out to at least 25$R_\sun$, far beyond the 3$R_\sun$ position of the remote wake in the U17 analysis.

The predicted signatures of the remote wake encounter are more mixed overall.  We still see a detectable increase in $V_r$, however, here it is only about 200 km s$^{-1}$ above the background Parker wind, rather than the roughly $2000$ km s$^{-1}$ enhancement seen in the leading front and the immediate wake.  There is no significant change in $B_r$ in this region and it is no longer strongly correlated to $V_r$, indicating that the plasma is now more compressional.  Correlated large deviations in $V_{\theta\phi}$ and  $B_{\theta\phi}$ are not present in this jet region, as mentioned earlier, however there is a small scale oscillation in the transverse components which is correlated strongly at the beginning of the encounter, though the correlation begins to fade near the end of the encounter.  $R_w$ varies during the encounter and is systematically below one, indicating that this region is no longer purely Alfv\'enic, and becomes even less so as later parts of the jet encounter the spacecraft.  There is an increase in $\rho$ concurrent with the enhanced $V_r$ inside the jet; however, this change is small compared to the density of the trailing dense-jet region.

\subsection{Case 3: Dense Jet}
\label{subsec:djregion}

\begin{figure*}
\begin{center}
\includegraphics[width=14.5cm]{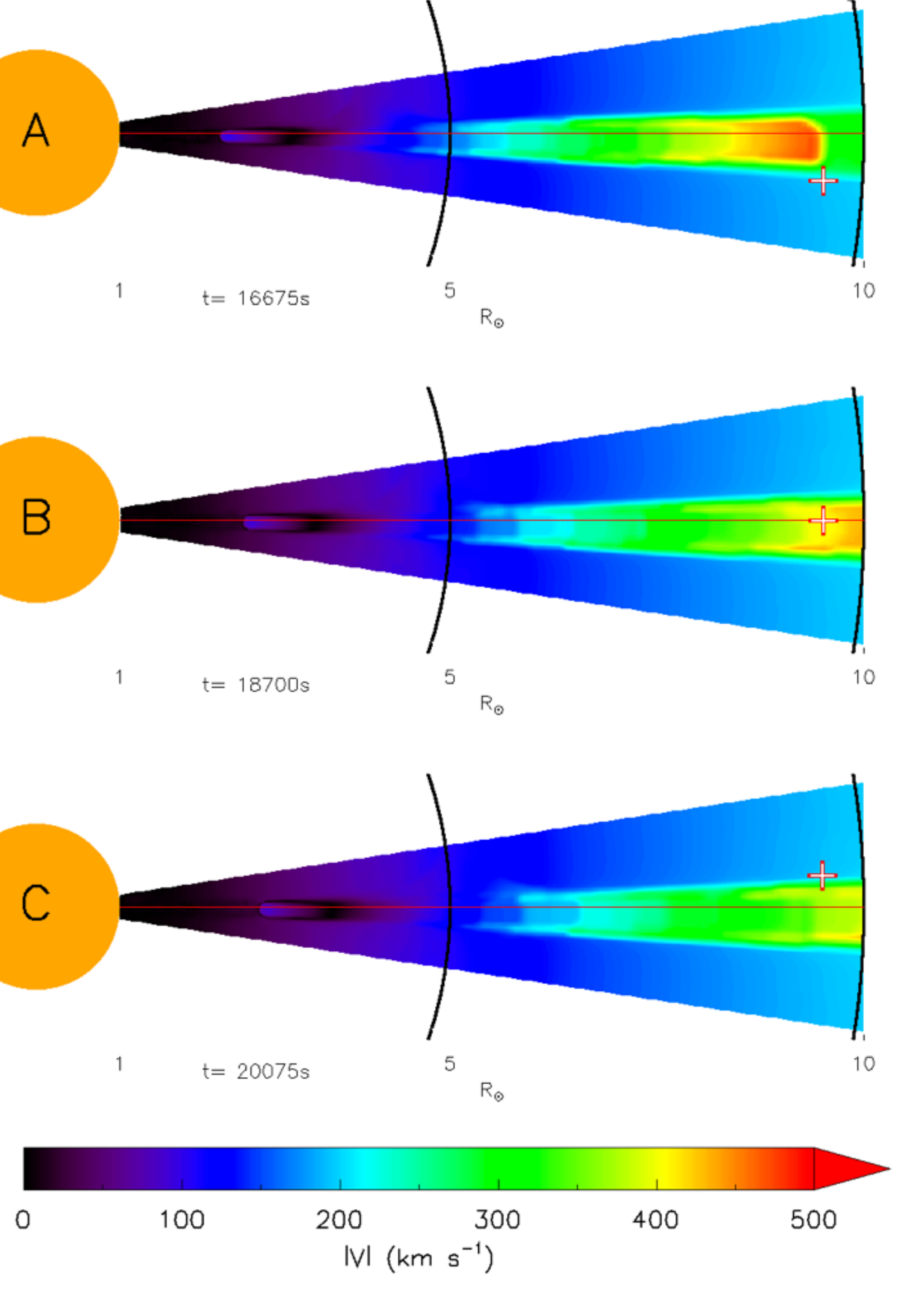}
\caption{Case 3: Same as Figure \ref{fig:imimages} for the dense jet fly-through. Each image (A, B, C) corresponds in time to the identically labeled dashed vertical line drawn in Figure \ref{fig:djplots}.}
\label{fig:djimages}
\end{center}
\end{figure*}

\begin{figure*}
\begin{center}
\includegraphics[width=17cm]{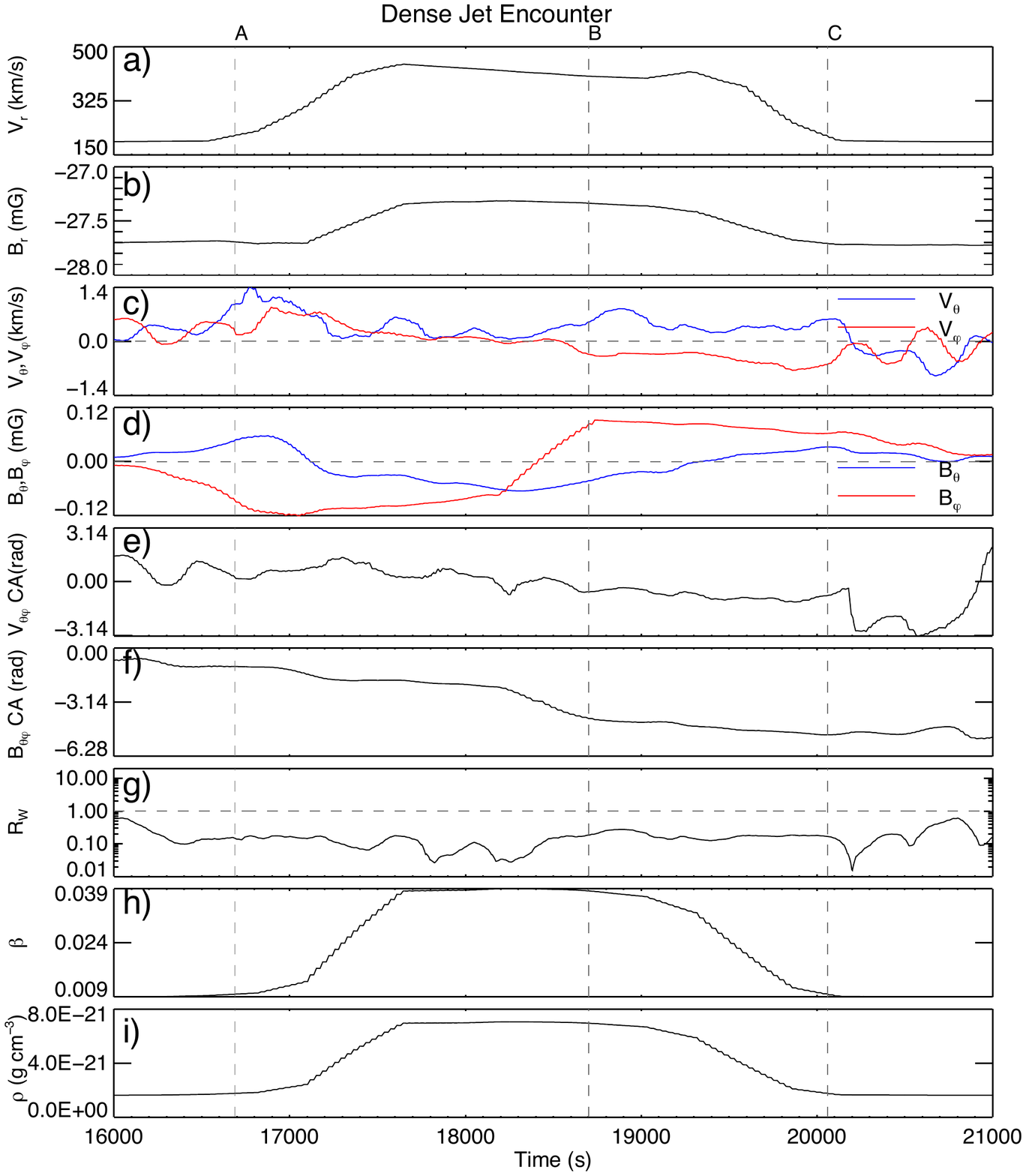}
\caption{Time series of plasma variables for the dense jet fly-through. Each dashed vertical line indicates a time at which an image is presented in Figure \ref{fig:djimages}.}
\label{fig:djplots}
\end{center}
\end{figure*}

The dense jet region is the part of the jet structure typically observed by remote-sensing EUV and X-ray instruments.  Figure \ref{fig:djimages} shows the spacecraft entering and then exiting the region of swift, dense jet plasma.  As in the immediate wake, the plasma is faster at the leading edge of the dense jet, but unlike the immediate wake, the dense jet shows a more homogeneous velocity structure inside the structure, without any of the complex and spatially compact features found in the shocked plasma of the immediate wake. The simulated PSP measurements are shown in Figure \ref{fig:djplots}.  

The simulated PSP spacecraft begins at coordinates $\left[9.54R_\sun,-4.5^{\circ},0^{\circ}\right]$ at time $t=15,400$ s, and follows the calculated trajectory.  PSP enters the jet at $t=16675$ s and exits the jet at $t=20075$ s, spending $3400$ s inside the jet, yielding $w_{eff} \approx 6.5 \times 10^5$ km = 0.92$R_\sun$ at this point.  In Figure \ref{fig:djplots}a, $V_r$ increases from a background value of 193 km s$^{-1}$ to a maximum of 443 km s$^{-1}$ within the jet.   This 250 km s$^{-1}$ enhancement in $V_r$ is about twice that seen in the remote wake (although only 10\% of the enhancement in the immediate wake). It is correlated in time with the density enhancement (Fig.\ \ref{fig:djplots}i), which is more than four times greater than the background solar wind density at this altitude and much larger than in the immediate or remote wakes.  $V_{\theta\phi}$ and $B_{\theta\phi}$ (Fig.\ \ref{fig:djplots}c,d) remain small and are completely uncorrelated with the radial magnetic and velocity fields (Fig.\ \ref{fig:djplots}a,b).  The large-scale oscillation in $B_{\phi}$ is still present, indicating (as suggested in \S \ref{subsec:imwake}) that $B_{\phi}$ varies persistently throughout the jet structure.  The $V_{\theta\phi}$ clock angle (Fig.\ \ref{fig:djplots}e) oscillates while undergoing a partial rotation during the encounter, whereas the $B_{\theta\phi}$ clock angle (Fig.\ \ref{fig:djplots}f) is smooth and nearly completes a full rotation.  $V_r$, $B_r$, $\beta$, and $\rho$ exhibit generally coordinated behavior (Fig.\ \ref{fig:djplots}a,b,h,i), with $V_r$ rising (and plateauing) slightly earlier and falling slightly later than the others.  The Wal\'en ratio (Fig.\ \ref{fig:djplots}g) is at least an order of magnitude less than one throughout the encounter. This, along with the large density increase, indicates that the spacecraft is in a more compressible plasma regime than in the immediate- and remote-wake regions.

Hence the most striking signature of the dense jet is the fourfold increase in mass density clearly observed as the spacecraft enters the radially outflowing plasma, simultaneous with an enhancement in the radial velocity.  As in the remote wake, the transverse components of both the velocity and magnetic field are small, and show no signs of cross-correlation.

\subsection{Case 4: Co-rotational Encounter}
\label{subsec:corot}

\begin{figure*}
\begin{center}
\includegraphics[width=14.5cm]{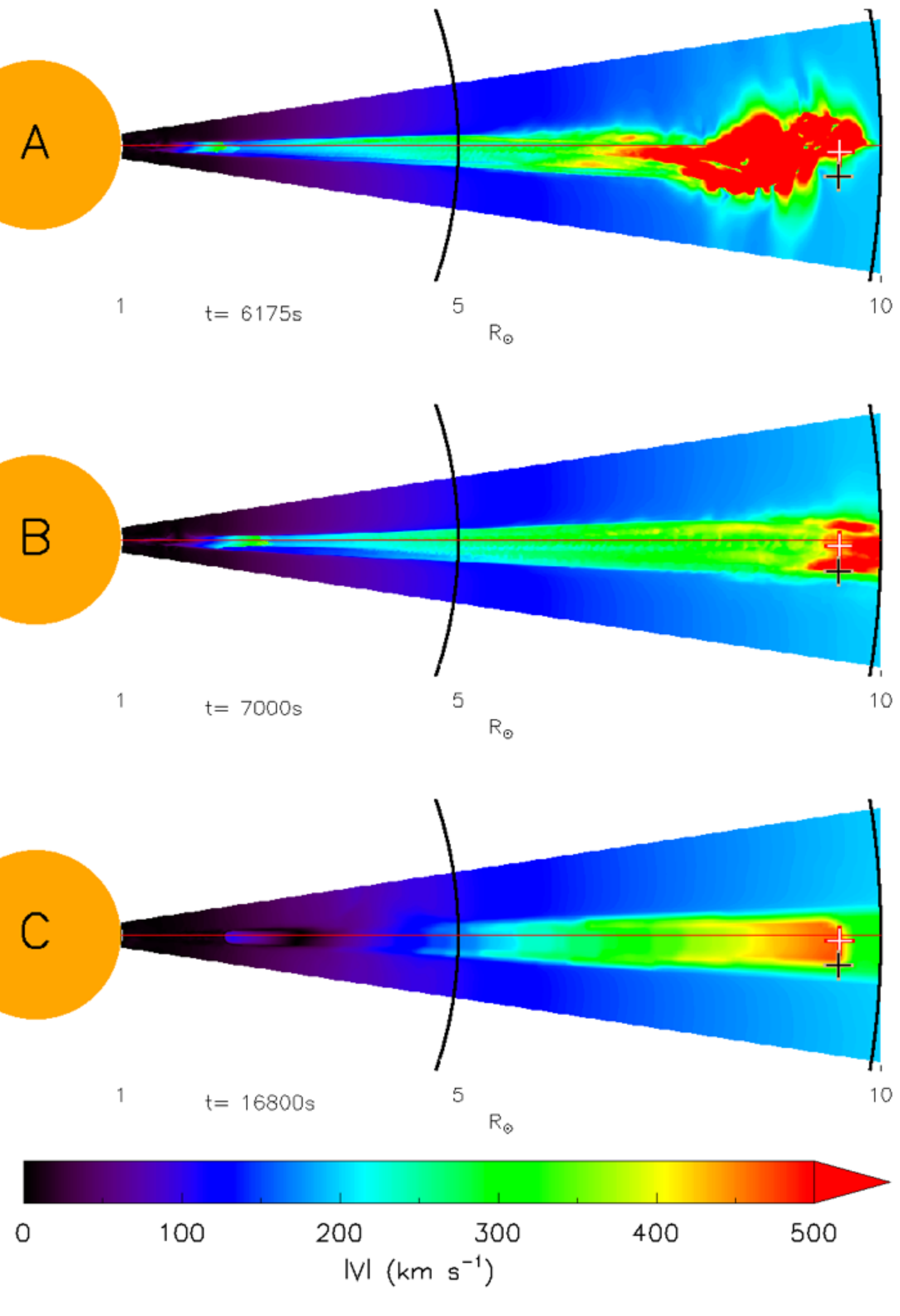}
\caption{Case 4: Same as Figure \ref{fig:imimages} for the co-rotational fly-through. Each image (A, B, C) corresponds in time to the identically labeled dashed vertical lines drawn in Figures \ref{fig:corotplots} and \ref{fig:offaxisplots} for the PSP spacecraft positioned at the red/white and black/brown crosses, respectively.}
\label{fig:corotimages}
\end{center}
\end{figure*}

\begin{figure*}
\begin{center}
\includegraphics[width=17cm]{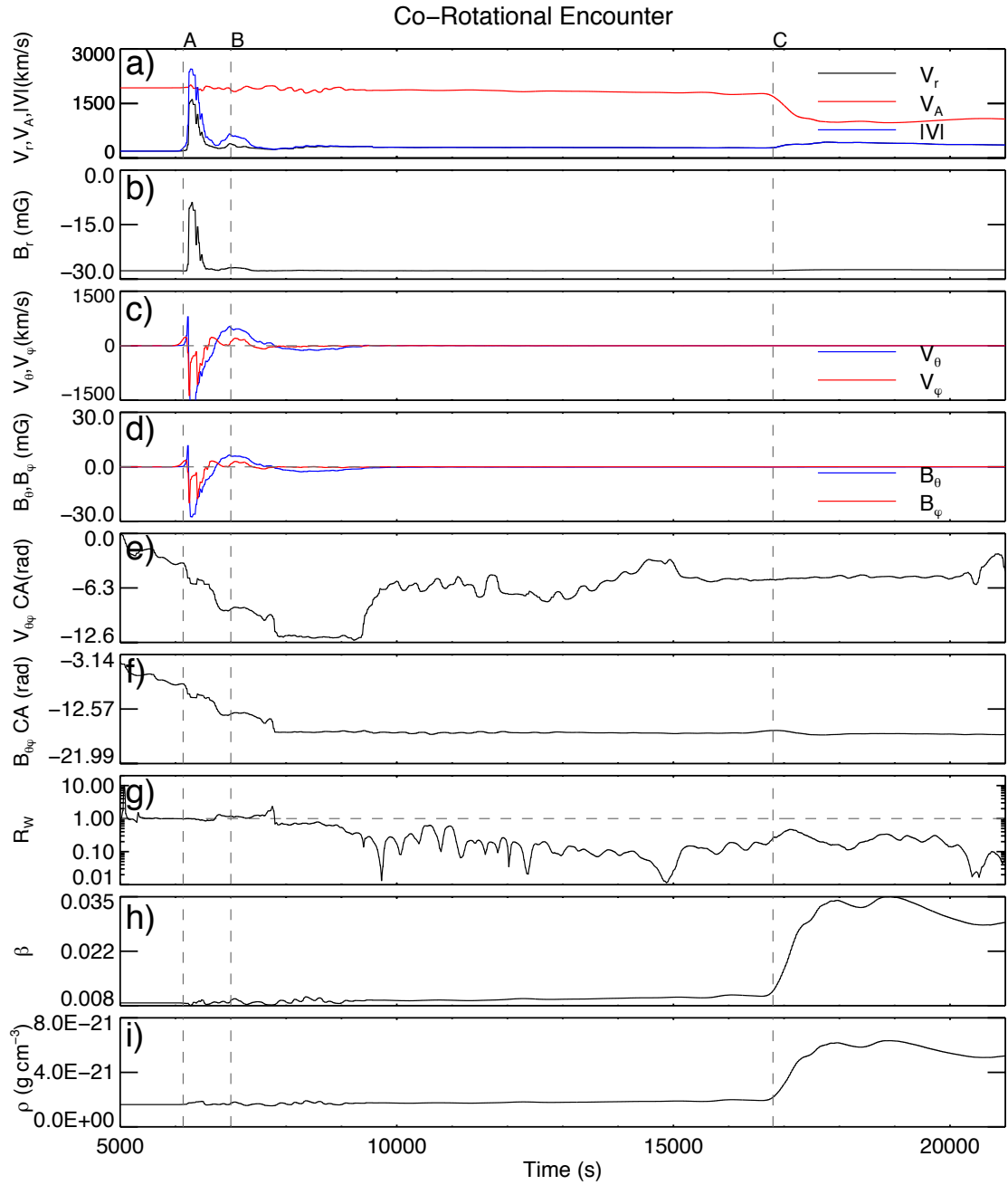}
\caption{Time series of plasma variables for the on-axis co-rotational fly-through. Each dashed vertical line indicates a time at which an image is presented in Figure \ref{fig:corotimages} for the PSP spacecraft position indicated with the red/white crosses.}
\label{fig:corotplots}
\end{center}
\end{figure*}

This run examines a possible jet encounter during one of PSP's co-rotational phases, where the spacecraft velocity is approximately equal to the rotational velocity of the solar surface, placing PSP temporarily in a heliostationary position.  As noted in \citet{fox16}, there will be two periods in each PSP orbit (one inbound and one outbound at $r \approx$ 35$R_\sun$) where the spacecraft will be approximately co-rotational with the Sun.  During these intervals PSP will perform fast radial scans, where the spacecraft will sample the solar wind over large radial distances while remaining over the same region of the solar surface.  In order to model this configuration, we allowed our simulated spacecraft to `hover' over the central region of the jet from $t=3900$ s to $t=21000$ s.  This simulated co-rotational phase is both radially closer and of longer duration than the corresponding expected PSP co-rotational phase. However, this radial scan through all three major dynamic regions of the CH jet serves as a baseline for identifying features of jet crossings in PSP observations that may be of shorter temporal duration and/or higher altitude.

Figure \ref{fig:corotimages} shows that all three of the previously discussed regions pass over the spacecraft location: the immediate wake in \ref{fig:corotimages}A, the remote wake in \ref{fig:corotimages}B, and the dense region in \ref{fig:corotimages}C.  Some features and trends inside the extended jet are more apparent from this perspective than in the previous encounters.  In Figure \ref{fig:corotplots}a,b the encounter with the high-velocity, reduced $B_r$ shock front at $t \approx 6100$ s is clear and correlates well with the immediate-wake encounter. An expected sudden drop in the Alfv\'{e}n speed appears when the dense region is encountered, and this change correlates with the large increase in $\rho$ found in Figure \ref{fig:corotimages}i.  Figure \ref{fig:corotplots} also confirms that, while the mass density fluctuates modestly in the shocked plasma of the immediate wake, the bulk of the plasma is located in the lower, dense-jet region, which encounters the spacecraft at $t \approx 16,800$ s.  Figures \ref{fig:corotplots}c,d show the sudden, correlated enhancement of $V_{\theta\phi}$ and $B_{\theta\phi}$ and their subsequent weakening as the shock front moves farther away from the spacecraft.   The $V_{\theta\phi}$ and $B_{\theta\phi}$ clock angles  (Fig.\ \ref{fig:corotplots}c,g) rotate fully in the immediate-wake region, then progress toward a unidirectional configuration with only slight variations after exiting this region. $R_w$ maintains a value of $+1$ during the entire encounter with the shock front and immediate-wake region, then deviates from this value throughout the rest of the jet (Fig.\ \ref{fig:corotplots}g). Thus, while the foremost part of the jet flow is incompressible and Alfv\'{e}nic in nature, its lower regions are not. It is interesting to note that this radial scan through the center of the jet reveals a ``hollow'' core where $\rho$ shows little enhancement, at least until an encounter with the lower-altitude dense region.  As a result, two of our three jet regions are difficult to detect remotely, reinforcing the need for \textit{in situ} measurements.

\begin{figure*}
\begin{center}
\includegraphics[width=17cm]{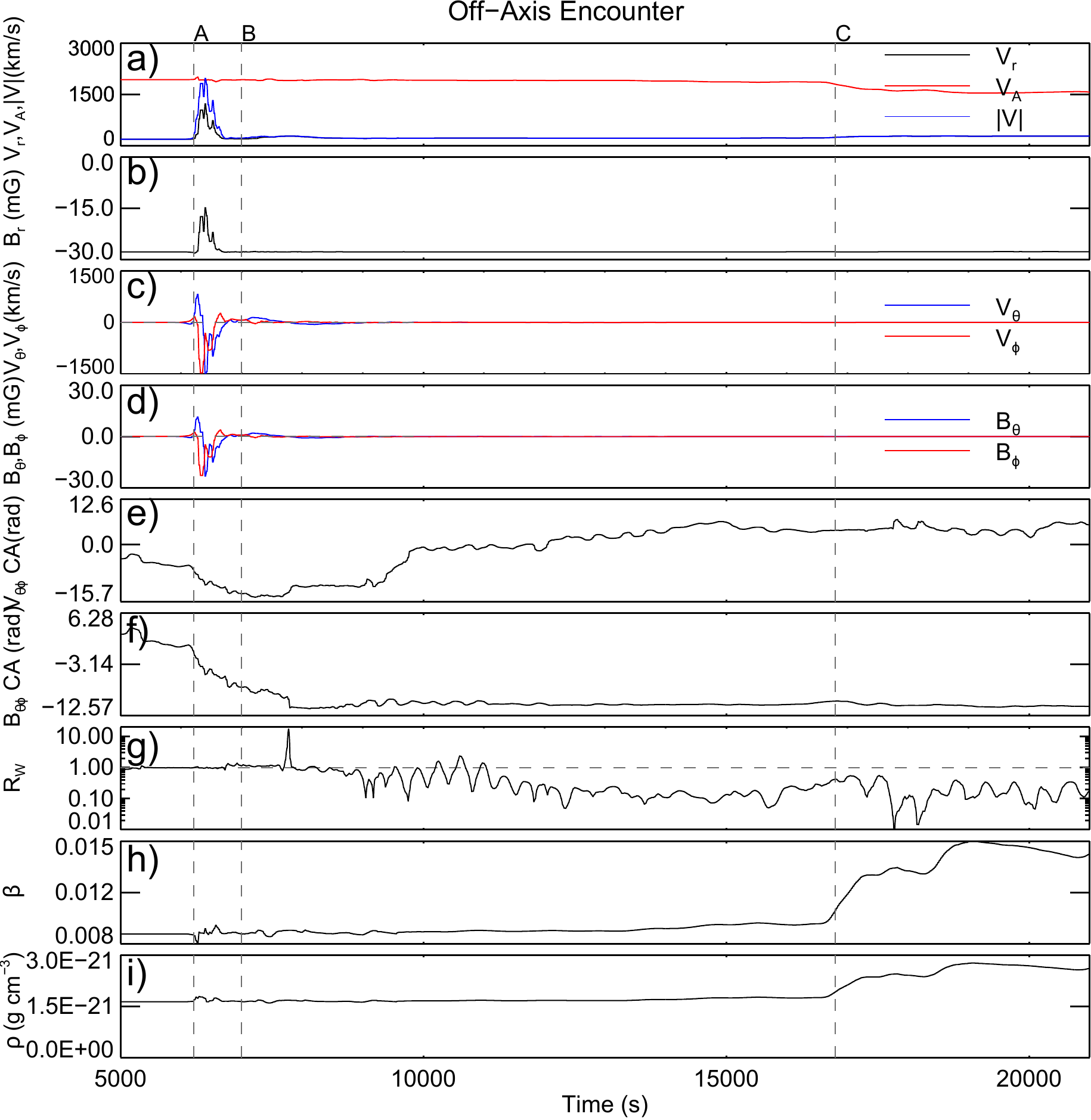}
\caption{Time series of plasma variables for the off-axis co-rotational fly-through. Each dashed vertical line indicates a time at which an image is presented in Figure \ref{fig:corotimages} for the PSP spacecraft position indicated with the black/brown crosses.  The scales in panels e, f, h, and i differ from those in Figure \ref{fig:corotplots}.}
\label{fig:offaxisplots}
\end{center}
\end{figure*}

It is unlikely that an actual PSP encounter will occur precisely at the jet axis.  To investigate the signatures of an off-axis encounter, we repeated the analysis just described, but shifted the simulated spacecraft location away from the centerline of the jet by 75\% of the jet radius.  As can be seen in Figure \ref{fig:offaxisplots}, some of the jet signatures are reduced in magnitude compared to the axial co-rotational encounter, yet they still stand out significantly from the background.

In this simulated encounter, decreases of about $40\%$ in the $V_r$ enhancement and $20\%$ in $V_{\theta\phi}$ (Fig.\ \ref{fig:offaxisplots}a), compared to the axial case, are observed, while roughly the same values of $B_{r}$, $R_{w}$, and $\rho$ are maintained in the immediate wake region (Fig.\ \ref{fig:offaxisplots}b,g,i).  In the dense jet, the off-axis encounter shows a roughly $55\%$ decrease in the maximum value of $\rho$, but this still represents a $150\%$ increase in $\rho$ over the background.  These results support our conclusion that even a significantly off-axis encounter should provide clearly discernible {\it{in situ}} signatures of CH jets at PSP.

As is seen most clearly in the co-rotational encounter, the internal plasma and field structures of the jet are generated at the reconnection site in the low corona, and then propagate outward after the jet launch.  Although the leading front outpaces the dense plasma region, the jet maintains its coherent structure as it propagates radially outward.  Our results suggest that an analysis of the jets encountered by PSP should reveal signatures persisting in the structure that illuminate the dynamics of the low-altitude regions that gave rise to the jet.

\section{Frequency of PSP/CH-Jet Encounters}
\label{sec:freq}

Although EUV and X-ray jets are relatively small in size and are transient, they occur quite frequently in coronal holes on the Sun. As noted in the Introduction and illustrated by Figure \ref{fig:wsjet}, some jets exhibit clear density signatures of passage outward through the corona in white-light coronagraphs. Moreover, a few \textit{in situ} observations of the solar wind have been interpreted as exhibiting velocity- and magnetic-field signatures of such jets or, at least, of the torsional Alfv\'en waves thought to be associated with them. In this section, we estimate the frequency of occurrence of PSP encounters with CH jets.

First, we estimate the frequency of occurrence of CH jets over the full Sun.  For simplicity, we assume that the total coronal-hole area, $A_{ch\odot}$, is roughly constant in time, and that at activity minimum it consists of the roughly 15\% of the total solar surface area that lies poleward of 60$^\circ$ latitude,
\begin{equation}
A_{ch\odot} \approx 0.15 \times 4\pi R_{\odot}^2 \approx 1 \times 10^{12} {\textup{ km}}^2.
\label{eq:achs}
\end{equation}
For the rate of occurrence, $S_{chj\odot}$, of CH jets per unit area, \citet{sako2013} analyzed Hinode X-ray observations to determine a value 
\begin{equation}
S_{chj\odot} \approx 1 \times 10^{-11} {\textup{ km}}^{-2} {\textup{ hr}}^{-1}.
\label{eq:schs}
\end{equation}
The total rate of occurrence, $N_{chj\odot}$, of CH jets over the Sun therefore is 
\begin{equation}
{
\begin{split}
N_{chj\odot} = S_{chj\odot} \times A_{ch\odot} &\approx 10 {\textup{ hr}}^{-1} \\
                                               &\approx 240 {\textup{ d}}^{-1}.
\end{split}
}
\label{eq:nchs}
\end{equation}
This global estimate agrees well with the 60 d$^{-1}$ rate of occurrence of CH jets measured by \citet{savcheva2007} in the Earth-facing South polar coronal hole, which accounts for 25\% of the global coronal hole area. For simplicity, we assume that the jets occur randomly with equal probability over the coronal-hole area, although \citet{sako2013} reported a tendency for higher rates of occurrence at the periphery. We point out here that this clustering may be irrelevant for \textit{in situ} analysis, since the open flux originating in coronal holes expands laterally to fill the entire heliosphere well within 10$R_{\odot}$, even though the flux occupies only a small fraction of the Sun's surface area.

Second, we estimate the fractional area of coronal holes that is occupied by the regions that give rise to jets.  \citet{sako2013} reported an average jet width
\begin{equation}
w_{chj} = 4 \times 10^3 {\textup{ km}},
\label{eq:wchj}
\end{equation}
whose corresponding surface area is
\begin{equation}
A_{chj} = \pi \left( \frac{w}{2} \right)^2 \approx 1 \times 10^7 {\textup{ km}}^{2}.
\label{eq:achj}
\end{equation}
We use this as a measure of the diameter of the jet source region on the surface.  The fractional coronal-hole area, $f_{chj}$, that is occupied by a typical jet source region then becomes
\begin{equation}
{f_{chj} = \frac{A_{chj}}{A_{ch\odot}} \approx 1 \times 10^{-5}.}
\label{eq:fchj}
\end{equation}
This is also the fractional area, on a sphere at the radial position of PSP, that is occupied by the footprint of the jet source region when it is mapped onto that sphere. We note that the distribution of jet widths reported by \citet{savcheva2007} (shown in their Figure 6) has an average value that is about 50\% larger, corresponding to about twice the surface area, compared to those quoted above.

Third, in order for PSP to detect a jet, the jet must pass within a circle that is centered on the spacecraft and whose radius is the projected width (diameter) of the jet. That is, the area of that circle is four times the area of an individual jet. Hence, the fractional area on the sphere sampled by PSP, $f_{PSP}$, is
\begin{equation}
f_{PSP} = 4 f_{chj} \approx 4 \times 10^{-5},
\label{eq:fpsp}
\end{equation}
and the corresponding footprint area on the solar surface, $A_{PSP}$, similarly is four times the jet footprint area,
\begin{equation}
A_{PSP} = 4 A_{chj} \approx 4 \times 10^7 {\textup{ km}}^{2}.
\label{eq:apsp}
\end{equation}
Finally, the expected rate of observations of jets over a coronal-hole region with that area is 
\begin{equation}
{
\begin{split}
N_{PSP} = S_{chj\odot} \times A_{PSP} &\approx 4 \times 10^{-4} {\textup{ hr}}^{-1} \\
								&\approx 1 \times 10^{-2} {\textup{ d}}^{-1}.
\end{split}
}
\label{eq:npsp}
\end{equation}
By this estimate, therefore, PSP should encounter an average CH jet about once in 100 days, based on the parameters of \citet{sako2013}. The expected frequency doubles to once in 50 days using the larger jet widths reported by \citet{savcheva2007}. Jets occur with a broad distribution of widths, and larger jets are more likely to be sampled by PSP proportional to the squares of their widths, i.e. to their areas. However, the greater ease of detecting the larger jets may be more than countered by their lower rate of occurrence: it is the product of area and occurrence rate that is determinative. Even at the lowest estimated frequency, PSP potentially will encounter some two dozen jets over the projected duration of the mission.

\section{Summary}
\label{sec:summary}
\begin{table*}
\begin{center}
\caption{Summary of Predicted \textit{In Situ} Jet Properties by Region}
\begin{tabular}{| c | c | }
\hline
{Region}				& {Jet Properties} \\
\hline
					& {Large-amplitude fluctuations in {\bf B} and {\bf V}} \\
{\it {Alfv\'{e}n wave front}}	& {Trans-Alfv\'{e}nic enhancements in $|V|$} \\
{\it {+}}				& {Rotations in {\bf B}$_{\theta\phi}$ and {\bf V}$_{\theta\phi}$ signaling helical structure} \\
{\it {immediate wake}}		& {Strongly coupled fluctuations in {\bf B}$_{\theta\phi}$ and {\bf V}$_{\theta\phi}$} \\
					& {Wal\'{e}n ratio $R_w \approx 1$}  \\
					& {Small ($\approx$ 20\%) increase in $\rho$} \\
\hline
					& {Small-amplitude fluctuations in {\bf B} and {\bf V}$_{\theta\phi}$} \\
					& {Trans-sonic enhancements in $V_r$} \\
{\it {remote wake}}		& {Weakly coupled fluctuations in {\bf B}$_{\theta\phi}$ and {\bf V}$_{\theta\phi}$} \\
					& {Wal\'{e}n ratio $R_w \approx 10\%$} \\
					& {Small ($\approx$ 10\%) increase in $\rho$} \\
\hline
					& {Small-amplitude fluctuations in {\bf B} and {\bf V}$_{\theta\phi}$} \\
					& {Trans-sonic enhancements in $V_r$} \\
{\it {dense jet}}			& {Decoupled fluctuations in {\bf B}$_{\theta\phi}$ and {\bf V}$_{\theta\phi}$} \\
					& {Wal\'{e}n ratio $R_w \approx 1\%$} \\
					& {Large ($\approx$ 300\%) increase in $\rho$} \\

\hline
\end{tabular}
\label{tab:jet_properties}
\end{center}
\end{table*}

{We have presented results from a simulated PSP encounter with a fully 3D MHD model of a solar coronal-hole jet.  Our simulation begins with a magnetic bipole embedded in a unipolar open and radial magnetic field, in a spherical, gravitationally stratified corona with an isothermal Parker solar wind.  The simulation introduces the magnetic free energy that powers the CH jet by slowly rotating the embedded minority-polarity intrusion. The stressed flux eventually undergoes a kink-like instability within the separatrix dome surrounding the closed magnetic field.  Reconnection through current patches on the separatrix drives a nonlinear, torsional Alfv\'{e}n wave, trailed by a more slowly moving spire of dense material, outward into the corona and inner heliosphere along the open magnetic field lines of the coronal hole.  Our simulation tracks the leading wave front to about 38$R_\sun$ and the trailing dense jet to about 13$R_\sun$.}

{Using the planned orbital parameters for the PSP mission, we constructed simulated spacecraft paths through the jet model and adjusted the start times so that the spacecraft would traverse each of the jet's three characteristic regions -- immediate wake, remote wake, and dense jet.  We also simulated a case where the spacecraft hovers directly above the jet source, in order to simulate the co-rotational period that PSP is planned to experience on either side of its perihelia.  By modeling these encounters, we have shown that PSP fly-throughs of coronal-hole jets will enable significant insights into the internal structure and physical processes of these ubiquitous impulsive events.  Furthermore, we have described specific, identifiable signatures of the jets produced by our reconnection-driven model that survive into the inner heliosphere, where they may be detected \textit{in situ} by PSP during its mission.  

{Given the observed frequency and wide spatial distribution of CH jets, we expect that the PSP mission will detect these jets on numerous occasions.  The observations are expected to exhibit different signatures, depending on which jet regions are traversed, as summarized in Table \ref{tab:jet_properties}.  The jet wave front is a switch-on Alfv\'{e}nic shock, manifested by a sudden increase of the velocity magnitude to approximately the local Alfv\'{e}n speed.  Within the immediate wake behind the front, all three components of the vector velocity and magnetic field undergo large-amplitude excursions.  Other signatures include rotations of the transverse velocity and magnetic field vectors over an angle of about $2\pi$, with strongly coupled fluctuations in the two quantities throughout the wake.  The Wal\'{e}n ratio should be very close to unity, indicating the incompressible, Alfv\'{e}nic nature of this region.  These distinctive signatures in {\bf B} and {\bf V} are accompanied by only small increases in the plasma density.}

{In the transitional remote wake of the jet, the fluctuations in the vector fields are much less pronounced, and have small amplitudes except for the radial (ambient) flow velocity.  The field-aligned flow exhibits trans-sonic enhancements associated with acoustic waves.  Hence, there is a mixture of Alfv\'{e}nic and compressional waves in this region, and the fluctuations in the transverse velocity and magnetic field are more weakly coupled than in the immediate wake.  The Wal\'{e}n ratio decreases to on the order of 10\% as the dominance of Alfv\'{e}n waves vs.\ acoustic waves subsides.  Increases in the plasma density remain small in the remote wake.}

{The prominent signature of the trailing dense-jet region is a very large density enhancement, $\approx $400\% in our simulated jet.  The compressional wave oscillations become dominant in this region as the Wal\'{e}n ratio declines further (to about 1\%), and the field-aligned flow component exhibits trans-sonic enhancements associated with the dense outflowing material.}

{All of the signatures listed above and tabulated in Table \ref{tab:jet_properties} are prominent in our simulation. It should be noted, however, that the simulation here described has a steady background solar wind prior to the launch of the jet.  Such idealized conditions cannot be expected to occur in the real solar wind, in general, so the weaker signatures exhibited by our jet may not be clearly distinguishable from background fluctuations.  Nonetheless, the stronger signatures of our jet are very distinctive, and should remain observable even in a more turbulent solar wind flow.  As we have shown, even significantly off-axis jet encounters preserve those readily discernible signatures.  Hence, we anticipate that they will stand out in sharp contrast with the ambient solar wind.

Our simulation results support the suggestion that the transient velocity enhancements detected by HELIOS at 0.3 AU could be direct manifestations of CH jets propagating through the heliosphere \citep{horbury2018} . However, Horbury et al. also assert that the plasma outflow in our simulated jets should not propagate to large distances but, instead, merge back into the ambient medium.  By the end of the simulation, the dense jet region propagates to at least 13$R_\odot$. This is not surprising, as the outflow is aligned with the open magnetic field and escapes in the wake of the leading Alfv\'{e}nic front. Therefore, in the absence of strong turbulence, we expect that even the dense plasma structure of our jet would be convected out of the lower corona relatively unperturbed across large heliocentric distances.

Our rough estimate of the frequency of encounters of PSP with CH jets suggests that at least two dozen such opportunities to obtain \textit{in situ} jet data should occur over the course of the mission. In future, it may be useful to refine both the interpolation of the PSP orbit and the time cadence of the model output in order to derive the predicted jet characteristics along the actual spacecraft track through a jet.  It is worth observing that the dynamics and structure of a CH jet, and especially of the Alfv\'enic shock front associated with it, should contain a multitude of kinetic-scale effects, such as temperature anisotropies, wave-particle interactions, plasma instabilities, and parallel electric fields, which can strongly influence jet evolution and which, by definition, are not captured by MHD.  Fortunately, PSP will be able to measure key physical properties at these kinetic scales, providing an unparalleled perspective into the true nature of these phenomena.  It would also be important to investigate the role of the photospheric driver using remote-sensing imaging data conjugate with the \textit{in situ} PSP measurements. Non-steady photospheric flows beyond the simplified boundary flow pattern incorporated in our model could significantly change the free energy input rate \citep{uritsky13} and modify the triggering conditions and the internal structure of the jet. Investigation of these effects should provide new insights into the physical mechanism of coronal hole jets emitted and propagating in a realistic solar environment.

\section{Acknowledgments}
The authors thank A Szabo for providing the projected PSP orbital coordinates used in this work. MAR also thanks S K Antiochos, S E Guidoni, P F Wyper, J T Dahlin, and M S Kirk for helpful conversations and feedback.  All of us appreciate helpful comments from our anonymous referee, including the suggestion to add \S \ref{sec:freq} on the frequency of PSP/CH-jet encounters.  MAR and VMU were supported by NASA grants NNG11PL10A 670.036 and 670.136 to CUA/IACS. CRD and JTK were supported by NASA LWS and H-SR grants to GSFC to investigate coronal jets and their heliospheric consequences.  The numerical simulations were performed on Discover, the NASA Center for Climate Simulation computing system, under a NASA High-End Computing award to CRD.

\bibliography{PSP_JET_PAPER_SUBMITTED.bib}
\end{document}